\documentclass{elsarticle}

\usepackage{lineno,hyperref}
\modulolinenumbers[5]

\usepackage{graphicx}
\usepackage{subcaption}
\usepackage{tabularx}
\usepackage{enumitem}
\usepackage{amssymb}
\usepackage{booktabs} 

\usepackage{algorithmic}
\usepackage[vlined, ruled, boxed,linesnumbered]{algorithm2e}
\usepackage{amsmath}

\let\today\relax
\makeatletter
\def\ps@pprintTitle{%
	\let\@oddhead\@empty
	\let\@evenhead\@empty
	\def\@oddfoot{\footnotesize\itshape
		{} \hfill\today}%
	\let\@evenfoot\@oddfoot
}
\makeatother

\SetKw{Break}{break}
\let\oldnl\nl
\newcommand\nonl{%
	\renewcommand{\nl}{\let\nl\oldnl}}

\usepackage{color}

\usepackage{etoolbox}
\patchcmd{\pprintMaketitle}
{\ifvoid\absbox\else\unvbox\absbox\par\vskip10pt\fi}
{\ifvoid\absbox\else\clearpage\unvbox\absbox\par\vskip30pt\fi}
{}{}
\patchcmd{\pprintMaketitle}
{\hrule\vskip12pt}
{}
{}{}
\patchcmd{\pprintMaketitle}
{\hrule\vskip12pt}
{}
{}{}
\appto{\pprintMaketitle}{\clearpage}

\begin{document}

\begin{frontmatter}

\title{SLA-Aware Multiple Migration Planning and Scheduling in SDN-NFV-enabled Clouds}

\author[unimelb]{TianZhang He\corref{mycorrespondingauthor}}

\cortext[mycorrespondingauthor]{Corresponding author.}
\ead{tianzhangh@student.unimelb.edu.au}

\author[monash]{Adel N. Toosi}
\ead{adel.n.toosi@monash.edu}

\author[unimelb]{Rajkumar Buyya}
\ead{rbuyya@unimelb.edu.au}

\address[unimelb]{Clouds Computing and Distributed Systems (CLOUDS) Laboratory, School of Computing and Information Systems,The University of Melbourne, Parkville, VIC 3010, Australia}
\address[monash]{Department of Software Systems and Cybersecurity, Faculty of Information Technology, Monash University, Clayton, VIC 3800, Australia}

\begin{abstract}
In Software-Defined Networking (SDN)-enabled cloud data centers, live migration is a key approach used for the reallocation of Virtual Machines (VMs) in cloud services and Virtual Network Functions (VNFs) in Service Function Chaining (SFC). Using live migration methods, cloud providers can address their dynamic resource management and fault tolerance objectives without interrupting the service of users. However, in cloud data centers, performing multiple live migrations in arbitrary order can lead to service degradation. Therefore, efficient migration planning is essential to reduce the impact of live migration overheads. In addition, to prevent Quality of Service (QoS) degradations and Service Level Agreement (SLA) violations, it is necessary to set priorities for different live migration requests with various urgency. In this paper, we propose SLAMIG, a set of algorithms that composes deadline-aware multiple migration grouping algorithm and on-line migration scheduling to determine the sequence of VM/VNF migrations. The experimental results show that our approach with reasonable algorithm runtime can efficiently reduce the number of deadline misses and has a good migration performance compared with the one-by-one scheduling and two state-of-the-art algorithms in terms of total migration time, average execution time, downtime, and transferred data. We also evaluate and analyze the impact of multiple migration planning and scheduling on QoS and energy consumption. 

\end{abstract}


\end{frontmatter}


\section{Introduction}
With the rapid adoption of cloud computing, the requirement of providing Quality of Service (QoS) guarantees is critical for cloud services, such as Web, big data, virtual reality, and scientific computing. For the benefit of cloud administrators, it is essential to prevent violations of Service Level Agreements (SLAs) and maintain QoS in heterogeneous environments. 
Therefore, there has been a notable focus on the quality, efficiency, accessibility, and robustness of cloud services. For instance, the latency of service function chaining (SFC)~\cite{halpern2015service} should be optimized to benefit both network service providers and end users. 

As one of the major virtualization technologies to host cloud services, Virtual Machine (VM) is used to provide computing and network resources hosted in cloud data centers. Live VM migration is one of the key technology to relocate VMs between physical machines without disrupting the accessibility of services \cite{clark2005live}. 
Therefore, as a dynamic resource management tool, live VM migration can realize various objectives in resource rescheduling of data centers, such as consolidation, load balancing, host overbooking, fault tolerance, scheduled maintenance, or even Virtual Network Functions (VNF) relocating from edge to cloud data centers due to the change of user location~\cite{cziva2018dynamic}.

Although researchers have been trying to achieve the objectives of dynamic resource rescheduling through live migration \cite{cziva2018dynamic,son2017sla}, few studies have focused on the impact of live migration overheads \cite{deshpande2017,mann2012,he2019performance} 
and the sequencing of multiple migrations \cite{ghorbani2012,bari2014cqncr,wang2017virtual}.
Due to the end of life of some VMs and the variance of workloads in cloud computing, dynamic resource management constantly generates many migration requests in optimization rounds. 
As a result, multiple migration requests need to be scheduled. For example, dynamic resource management policies for performance efficiency and energy-saving \cite{beloglazov2012optimal} can generated up to 12500 migrations within 10 days. Moreover, commercial cloud platforms provides live migration to keep VM instances running during the host event, such as hardware or software update. For example, in Google Cloud Compute Engine, live migration occurs to one VM at least once every two weeks due to the software or hardware update\footnote{Google Cloud Compute Engine. \url{https://cloud.google.com/compute/docs/instances/setting-instance-scheduling-options}}. In 2020, to make the compute infrastructure cost effective, reliable and performant, Google Cloud Compute Engine also introduced dynamic resource management for E2 VMs through performance-aware live migration\footnote{Dynamic resource management in E2 VMs. \url{https://cloud.google.com/blog/products/compute/understanding-dynamic-resource-management-in-e2-vms}}.
Therefore, it is important to determine the order (sequence) of the migration tasks to optimize the total migration time \cite{ghorbani2012,wang2017virtual,bari2014cqncr}, which is the interval between the start of the first migration and the completion of the last migration.

In cloud data centers, Software-Defined Networking (SDN) can enable the centralized control of network resources in terms of network topology, connectivity, flow routing, and allocated bandwidth. The Virtual Network Functions (VNFs) hosted in cloud data centers can also be linked as a Service Function Chaining (SFC) \cite{halpern2015service} by SDN controller. Migration planning for VNFs in the chain is not trivial since SFC requires traffic to traverse through a certain sequence of VNFs. In addition, because migrations share the network resources with other services, it is necessary to efficiently plan and schedule migration tasks to reduce the overhead impact on the QoS of other applications. The migration planner and scheduler based on the SDN controller can manage the network resources in a fine-grained manner for the migration tasks and application services in terms of network routing and bandwidth allocation.

\textbf{Connectivity and Sequence:} Compared with services such as scientific computing, the connectivity and corresponding network requirement of links in SFC between source and destination are dynamically changing. This will also cause the remaining bandwidth of the migration to change. 
Furthermore, as the available bandwidth of the physical link changes according to the connectivity of the SFC, we also need to carefully consider the impact of the new placement of the VNF. 
As a result, the new placement will affect the rest of the migration requests that use the same path.
In addition, two migrations could be performed concurrently if there are no shared paths between them. However, performing multiple live migrations in arbitrary order will result in service quality degradation. Therefore, efficient planning of the migration sequence is crucial to reduce the impact of live migration overheads.

In addition to the optimization of total migration, several other parameters that affect migration performance are largely neglected:

\textbf{Scheduling window (deadline):} For migration such as scheduled maintenance, disaster evacuation, load balancing policy, and other dynamic allocation algorithms \cite{cziva2018dynamic, tsakalozos2017live}, it is usually associated with a time window (defined deadline) that requires the VM or VNF to be evacuated from the source and run on the destination host. For instance, the deadlines for SLA-related migration tasks are based on the violation speed and the cumulative violations threshold of Service Level Objective (SLO), such as response time and end-to-end delay. The scheduling window refers to the time interval between the arrival of migration task request and the deadline for the new placement. Failure to meet the deadline will result in QoS degradation and SLA violation. 

\textbf{Allocated bandwidth:} 
During the live VM migration, the applications running inside VM constantly modify the local stack and variables in the memory. The memory modified during the last round of dirty memory transmission needs to be transferred again. The goal of live migration is to reduce the memory difference between the source and destination hosts in order to stop the VM and copy the remaining dirty memory pages to the destination. A smaller memory difference in the stop-and-copy round means that the service has much shorter downtime. 
Therefore, live migration is highly dependent on the network for dirty memory transmission. We can consider live migration as a non-preemptive task. If the available bandwidth is lower than the rate of memory dirtying during the iterative transmission, then the data transferred previously used for migration convergence will be in vain. Furthermore,
although the average bandwidth for the entire process of a migration might be the same, the insufficient bandwidth at the beginning may severely extend the migration time. Therefore, we should carefully allocate available paths to multiple migration requests.

To help cloud providers guarantee QoS and SLAs during the multiple live migrations,
in this paper, we investigate the problem of optimizing the total migration time, transferred data, downtime, and average execution time of multiple VM/VNF migrations within the scheduling window in software-defined cloud data centers. We propose SLAMIG (SLA-aware Migration), which is a set of algorithms that includes the deadline-aware concurrent migration grouping for multiple VM/VNF migrations and an on-line migration scheduler to minimize the total migration time by considering the migration deadlines. 

The main \textbf{contributions} of this paper are summarized as follows: 
\begin{itemize}
	\item We modeled the multiple migration planning problem to optimize total migration time and deadline missing in the context of VMs/VNFs connectivity.
	\item We are the first to introduce the scheduling window for multiple migration scheduling.
	\item We investigated the impact of allocated bandwidth on the beginning of one migration.
	\item By maximizing the concurrent migration groups with minimal weight, we proposed a heuristic algorithm that achieves good performance for multiple migrations by considering the migration deadline.
	\item We designed an on-line migration scheduler to dynamically schedule migrations from different migration groups. 
	\item We not only analyzed the multiple migration scheduling in total migration time and downtime, but the average execution time, total transferred data, deadline violations, QoS, and energy consumption.
\end{itemize}

The rest of the paper is organized as follows. Section 2 introduces the system overview and background of the live migration. 
In Section 3, we present the motivation example, the impact of migration bandwidth, the model of sequential/parallel migrations, deadline of the migration, and the problem formulation of multiple migration planning. 
The summary of observation, rationales of algorithm design, and the details of proposed algorithms are presented in Section 4. In Section 5, experiment design and results are shown. Finally, we review the related work in Section 6 and conclude the paper in Section 7.

\section{System Overview and Background}
In this section, we first discuss the system overview (Fig. \ref{fig: system-overview}) and then we present the mathematical model of single live migration. 

\subsection{System overview}
\begin{figure}[t]
	\centering
	\includegraphics[width=\linewidth]{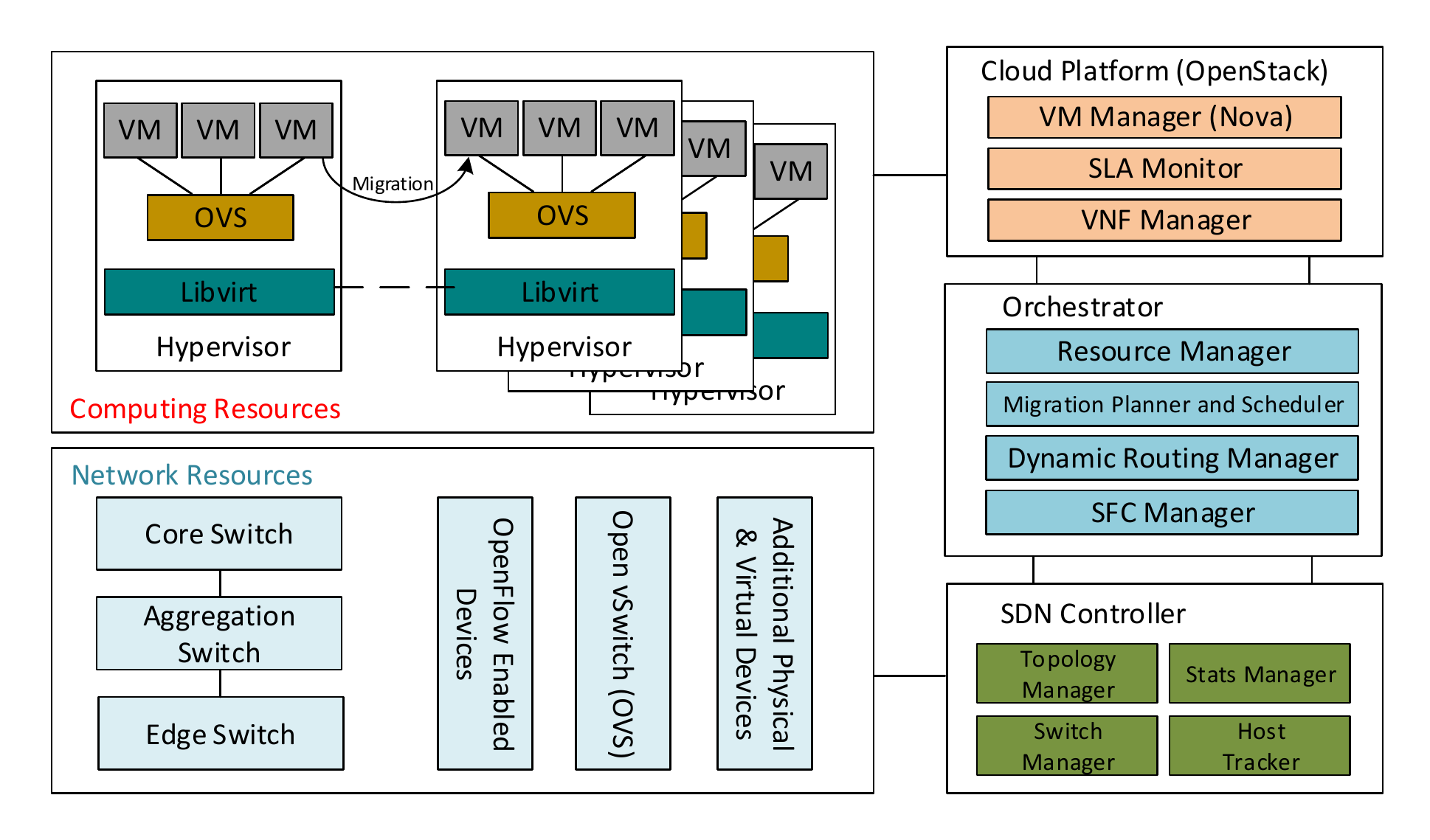}
	\caption{System overview}
	\label{fig: system-overview}
\end{figure}

In data centers, physical machines host VMs/VNFs to provide various cloud services. By utilizing the virtualization, we can separate the network functions (such as Intrusion Detection Systems (IDS), and Firewalls (FW)), web applications, and scientific computing servers from underlying computing and network resources. Thus, VNFs and VMs could be migrated from one physical host to another through live migration without disruption. To alleviate SLA violations and meet the QoS guarantees in the dynamic environment, the resource management policies \cite{cziva2018dynamic,son2017sla} or migration selectors \cite{mann2012,xu2014} determine which, when and where a VNF or VM should be migrated by predicting the future system state based on the historical data and current state of computing resources (physical hosts and virtual instances) and network resources (loads of links). 

The instance provision or VM manager (such as OpenStack Nova) controls the computing resources, while the SDN controller manages the network resources. By merging Software Defined Networking (SDN) \cite{son2017taxonomy} into cloud computing, the SDN-enabled cloud data centers provide a comprehensive and flexible solution to the increasing end-user applications and the dynamic network environment needed by SFC. By separating the data and control plane of network routing, the SDN controller can dynamically allocate bandwidth to services and control the routing of each network flow. With its centralized controller, it also provides a global view of the network topology, switch status, profiling, and monitoring the link statistics (bandwidth and latency), as well as dynamic connection management of different cloud services. 
For example, it can monitor on-going flows, calculate the `best' paths for migration requests and other cloud services, and dynamically connect VNFs and VMs to SFCs and Virtual Data Centers (VDCs) by pushing the forwarding rules through the OpenFlow protocol \cite{mckeown2008openflow}.	
On the other hand, the VM and VNF managers are responsible for configuring and assigning all computing and storage resources, and managing the lifecycle of VNF and VM. It also monitors the status of each VNF and VM, such as the dirty page rate and the speed of SLO violations.

Based on the centralized information of both network and computing resources, the orchestrator can calculate the optimal allocation of all VNFs and VMs. According to the optimal allocation, it generates multiple migration requests. Then, the migration planner needs to decide the sequence of multiple migrations considering the contention on the shared resources between migrations, such as network path and bandwidth. This can be realized by using an SDN controller which provides the network topology, dynamic network routing and bandwidth allocation. Then, the migration scheduler schedules migration tasks either sequentially or concurrently based on the shared resources. 

Our proposed system framework assumes that there is a queue for migration requests. Each migration task is defined by the following items: (1) the migrating VM/VNF; (2) the source-destination host pair and corresponding network routings; (3) the scheduling window (deadline). As shown in Figure \ref{fig: system-overview}, our proposed approach SLAMIG includes three components: (1) the migration planning module; (2) the on-line migration scheduler; and (3) the dynamic routing module. The objective of our approach is to plan and schedule migration tasks so that all migration tasks are completed within the deadline while meeting the SLA, thereby minimizing the impact of multiple migrations.

\subsection{Mathematical Model of Live Migration}\label{sec:single-model}
In order to better understand the impact of multiple migrations and the scheduling problem, we first need to establish the mathematical model of single live migration.
Live migration can be categorized into the pre-copy \cite{clark2005live} and post-copy \cite{shribman2012} memory migration. In the pre-copy live migration, the virtual machine monitor (VMM) iteratively copies dirty memory pages of VM/VNF running on the source host to the destination host. 
Since the pre-copy migration is the most commonly used technology for hypervisors (KVM, Xen, etc.), we consider it as the base model used in the multiple migration planning. 

\begin{figure}[t]
	\centering
	\includegraphics[width=\linewidth]{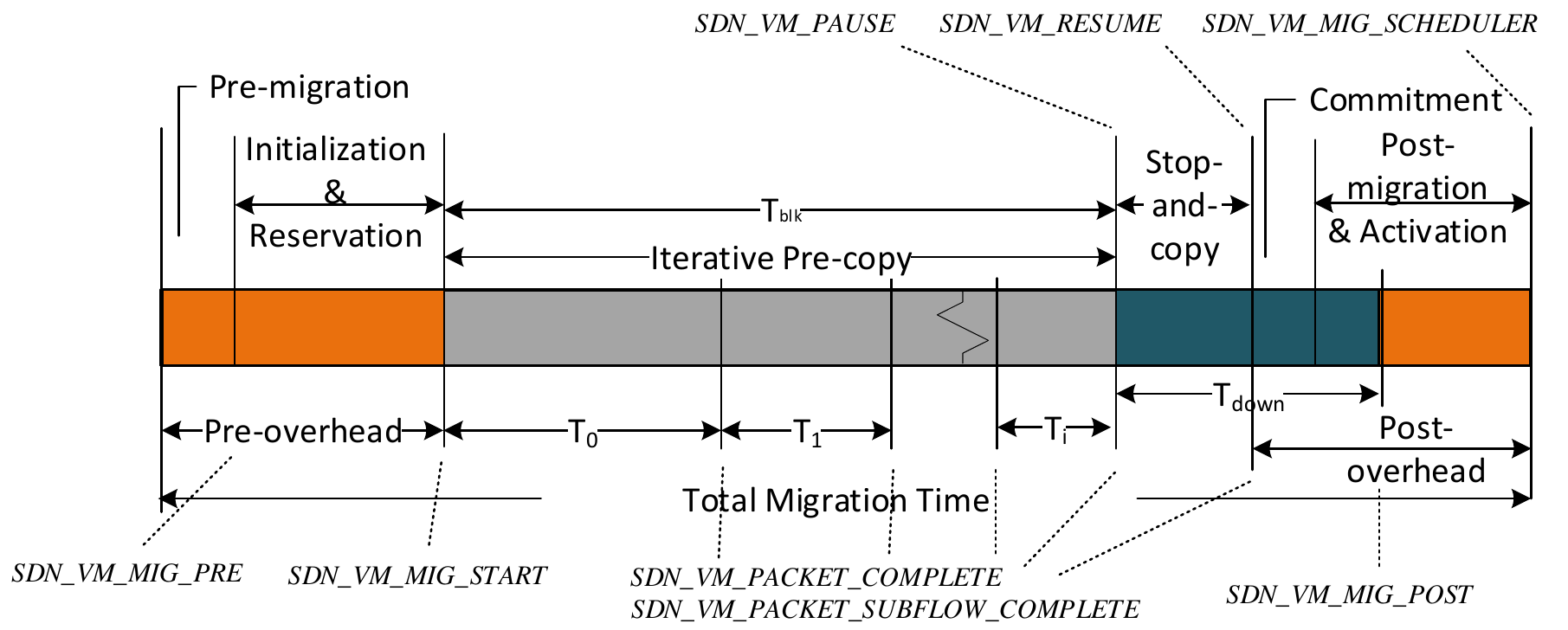}
	\caption{Pre-copy Live Migration}
	\label{fig: migration}
\end{figure}

According to the live migration process \cite{clark2005live}, the pre-copy live migration includes 8 steps, as shown in Fig. \ref{fig: migration}: 
Pre-migration, Initialization, Reservation, Iterative memory copy, stop-and-copy, Commitment, and Post-migration. 
Therefore, we can classify the three parts of migration overhead as computation (pre/post-migration) and network-related (iterative network transmission). The migration time or execution time refers to the time interval from the initialization of the pre-migration process on the source host to the successful completion of the post-migration process on the destination host \cite{he2019performance}. Therefore, the total single migration time $T_{mig}$ (also known as execution time in multiple migration scenario $T_{exe}$) can be represented as: 
\begin{equation}\label{eq:total-com-mem}
T_{mig} = T_{pre} + T_{mem} + T_{post}
\end{equation}

When $T_{mem}$ represents the iterative memory copy and stop-and-copy phases, the components of pre-migration and post-migration overheads can be expressed as:
\begin{equation}
\begin{array}{l}
{T_{pre}} = {\rm{PreMig}} + {\rm{Init}} + {\rm{Reserv}}\\
{T_{post}} = {\rm{Commit}} + {\rm{Act}} + {\rm{PostMig}}
\end{array}
\end{equation}

In the pre-copy live migration, the memory of VM is transferred over the network when the VM continues to operate \cite{clark2005live}. Thus, the modified memory during the previous iteration of memory copying needs to be transferred again.  There will be a small number of hotspot pages with the high frequency of updates which will be transferred within the stop-and-copy phase.
The downtime refers to the time interval between the VM suspension due to the stop-and-copy, commitment, and activation phases, as shown in Figure \ref{fig: migration}. From the user's perspective, the service is unavailable during the downtime. The dirty page rate is the rate of page dirtying per second traced by the page table of the hypervisor, e.g. Xen and KVM.  Since the behavior of the page dirtying is known at every point in time, the total transferred data in the stop-and-copy phase can be estimated and hence the downtime \cite{clark2005live}. Furthermore, in order to prevent a large downtime and iteration round due to a high page dirty rate compared to the available bandwidth, we need to set a threshold (upper-bound) for the downtime and the total number of allowed iteration rounds in practice.

We use $M$ to denote the memory size of VNF/VM, and let $\rho$ represent the average compression rate used in memory compression algorithm \cite{svard2011evaluation}. Let $R_i$ and $L_i$ denote the average dirty page rate and bandwidth in iteration round $i$. In total $n$ rounds of memory copying, $T_i$ denotes the time interval of $i_{th}$ iteration. As shown in Fig. \ref{fig: migration}, the transferred volume $V_i$ of $i_{th}$ round can be calculated as: 
\begin{equation}\label{eq: vi}
{V_i} = \left\{ {\begin{array}{*{20}{c}}
	{\rho  \cdot M}\\
	{\rho  \cdot {T_{i - 1}} \cdot {R_{i - 1}}}
	\end{array}\begin{array}{*{20}{c}}
	{{\rm{if\enspace}} i = 0}\\
	{{\rm{    }}otherwise}
	\end{array}} \right.
\end{equation}
where the $\rho$ is the percentage number of average compression rate, and the unit of $M$ is bits, $T_{i-1}$ is seconds (s), and $R_{i-1}$ and $L_i$ are bits per seconds (bps).

Based on Equation (\ref{eq: vi}), the transmission time of first round ($i = 0$) can be calculated as ${T_0} = {{\rho  \cdot M} \mathord{\left/
		{\vphantom {{\rho  \cdot M} {{L_0}}}} \right.
		\kern-\nulldelimiterspace} {{L_0}}}$. By submitting Equation (\ref{eq: vi}) into the result of first round, we get ${T_1} = {{\rho  \cdot {T_0} \cdot {R_0}} \mathord{\left/
		{\vphantom {{\rho  \cdot {T_0} \cdot {R_0}} {{L_1}}}} \right.
		\kern-\nulldelimiterspace} {{L_1}}} = {{{\rho ^2} \cdot M \cdot {R_0}} \mathord{\left/
		{\vphantom {{{\rho ^2} \cdot M \cdot {R_0}} {{L_0} \cdot {L_1}}}} \right.
		\kern-\nulldelimiterspace} {{L_0} \cdot {L_1}}}$.
	
Thus, the transmission time of $i_{th}$ round $T_i$ can be represented as:

\begin{equation}\label{eq: var-mem-copy}
{T_i} = {{{V_i}} \mathord{\left/
		{\vphantom {{{V_i}} {{L_i} = }}} \right.
		\kern-\nulldelimiterspace} {{L_i} = }}{\rho ^{i + 1}} \cdot {{\prod\nolimits_{j = 0}^{i - 1} {{R_j} \cdot M} } \mathord{\left/
		{\vphantom {{\prod\nolimits_{j = 0}^{i - 1} {{R_j} \cdot M} } {\prod\nolimits_{j = 0}^i {{L_j}} }}} \right.
		\kern-\nulldelimiterspace} {\prod\nolimits_{j = 0}^i {{L_j}} }}
\end{equation}

When $R_i$ and $L_i$ are constant, for the convergence migration, the average dirty page rate must be no more than the bandwidth in every iterations $0 \le \sigma  < 1$. 
Let ratio
$\sigma  = \rho  \cdot {R \mathord{\left/ {\vphantom {R L}} \right. \kern-\nulldelimiterspace} L}$. 
Then, ${T_i} = {{\rho  \cdot M \cdot {\sigma ^i}} \mathord{\left/
		{\vphantom {{\rho  \cdot M \cdot {\sigma ^i}} L}} \right.
		\kern-\nulldelimiterspace} L}$.	
Therefore, the total memory copying time $T_{mem}$ is:
\begin{equation}
\label{eq:mem-copy}
{T_{mem}} = \frac{{\rho  \cdot M}}{L} \cdot \frac{{1 - {\sigma ^{n + 1}}}}{{1 - \sigma }}
\end{equation}

Let $\Theta$ denote the maximum allowed number of iteration rounds and $T_{dthd}$ denotes the downtime threshold. Then, $V_{thd} = T_{dthd} * L_{n-1}$ as remaining dirty page need to be transferred in the stop-and-copy phase can be calculated.
We can calculate the total iterations $n$ by using the inequality $V_n \le V_{thd}$ in Equation \ref{eq:mem-copy}:
\begin{equation}\label{eq:round-number}
n = \min \left( {\left\lceil {{{\log }_\sigma }\frac{{{V_{thd}}}}{M}} \right\rceil ,\Theta } \right)
\end{equation}

Therefore, by using $T_{post}^{'}$ as the reassignment time of computing resources and network ports for the migrated instance, the actual downtime is represented as:
\begin{equation}
T_{down} = T_d + T_{post}^{'} < T_{thd} + T_{post}
\end{equation}

\section{System Model and Problem Formulation}
\subsection{Motivation Example}
\begin{table}[t]
	\caption{Configurations of different flavors}
	\resizebox{\linewidth}{!}{
		\begin{tabular}{|l|l|l|l|l|l|l|l|l|l|}
			\hline
			No. & Flavor & Mem & CPU & Disk	& No. & Flavor & Mem & CPU & Disk \\
			&		& (GB)&(cores)& (GB) 	&		&	& (GB)&(cores)& (GB)\\
			
			\hline
			1 & xlarge & 64 & 12 & 120 & 5 & tiny   & 2	& 1 & 2\\
			2 & large  & 16 & 8  & 60 & 6 & micro  & 1  & 1 & 1\\
			3 & medium & 8  & 4  & 20 & 7 & lb/ids/fw	& 8	& 10/12/16 & 8 \\
			4 & small  & 4	& 2	& 10 & 	8 & web/app/db	& 256 & 8/4/12 & 1000 \\
			\hline
		\end{tabular}	
	}
	\label{tb: flavor}
\end{table}	

\begin{figure}[t]
	\centering
	\vspace{-0.5cm}
	\includegraphics[width = 0.7\linewidth]{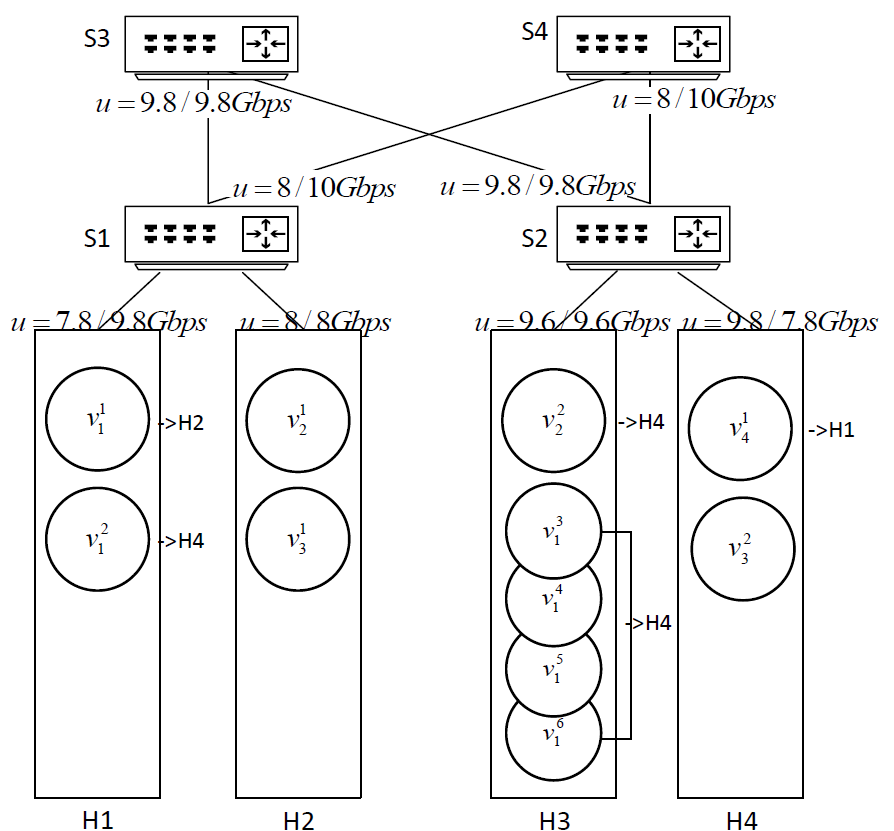}
	\caption{Initial mapping for VM/VNF $v^j_i$ of VDC/SFC $G^j$, migration requests $(s_j, d_j)$, and available bandwidth $u$ of upload/download interfaces}
	\label{fig: case-study}
\end{figure}
\begin{figure}[t]
	\centering
	\vspace{-0.2cm}
	\includegraphics[width = 0.7\linewidth]{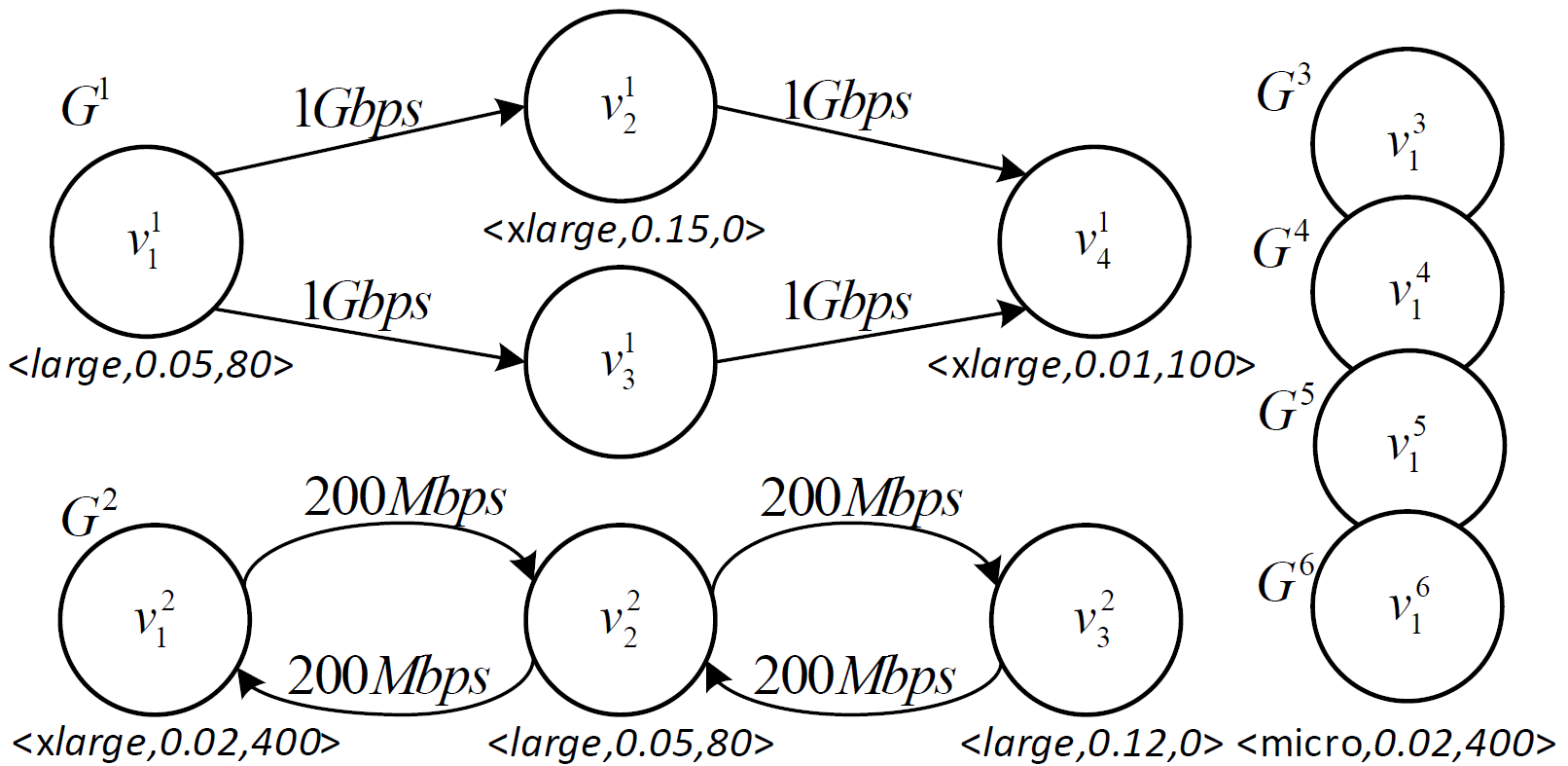}
	\caption{SFC and VDC configurations of $<$flavor, dirty page rate, migration deadline$>$}
	
	\label{fig: case-study-sfc}
\end{figure}

In this section, we discuss the problem and our motivation using an example of optimizing the total migration time to show the impact of migration orders on the total migration time, VNF/VM downtime, and SFC/VDC migration time and downtime.

Migration processes produce elephant flows which take a disproportionate part of network resources for a long time. At the end of each migration, the network flows within the data center network are redistributed accordingly due to the relocation of VNFs or VMs and their connectivity. With the change of available bandwidth in Data Center Network (DCN), the result of one migration will affect subsequent migrations that share the links with the completed one. The objective of migration planning is to find the orders of migrations to optimize the total live migration time of all requested migrations with certain constraints, such as the priority, required bandwidth, residual bandwidth on the links, and capacity of CPU, memory, and storage resources. 

In the network of tree topology shown in Fig. \ref{fig: case-study}, there are 4 switches which include 2 top-of-rack (S1 and S2), 2 aggregation switches (S3 and S4), and 4 hosts (H1 to H4). All the hosts and switches are connected through 10Gbps links. One SFC $G^1$, one VDC $G^2$ and four other VMs ($v_1^3$ to $v_1^6$) are hosted in different hosts accordingly. Fig. \ref{fig: case-study-sfc} shows the connectivity and reserved bandwidth of virtual links among instance with different flavors (Table \ref{tb: flavor}) of $G^1$ and $G^2$, as well as the dirty page rate and CPU, memory, and storage requirements. SFC $G^1$ contains 4 VNFs where $v_2^1$ and $v_3^1$ are the same type of VNF.
The migration time is composed of the processing time of pre-migration and post-migration overheads on computing resources and the network transmission time of the dirty memory. We assume the processing time of pre-migration and post-migration overheads on the single core are 0.8 seconds and 1.2 seconds, respectively. 

Fig. \ref{fig: case-study} illustrates the initial mapping of these VNFs/VMs and migration requests for another possible mapping in the physical topology. Let $u$ denote the residual bandwidth on the links. According to the reserved requirements of virtual links, we calculate the initial available inbound and outbound bandwidth of each network interface. 

\begin{figure}[t]
	\centering
	\begin{subfigure}[t]{.9\textwidth}
		\centering
		\includegraphics[width=\textwidth]{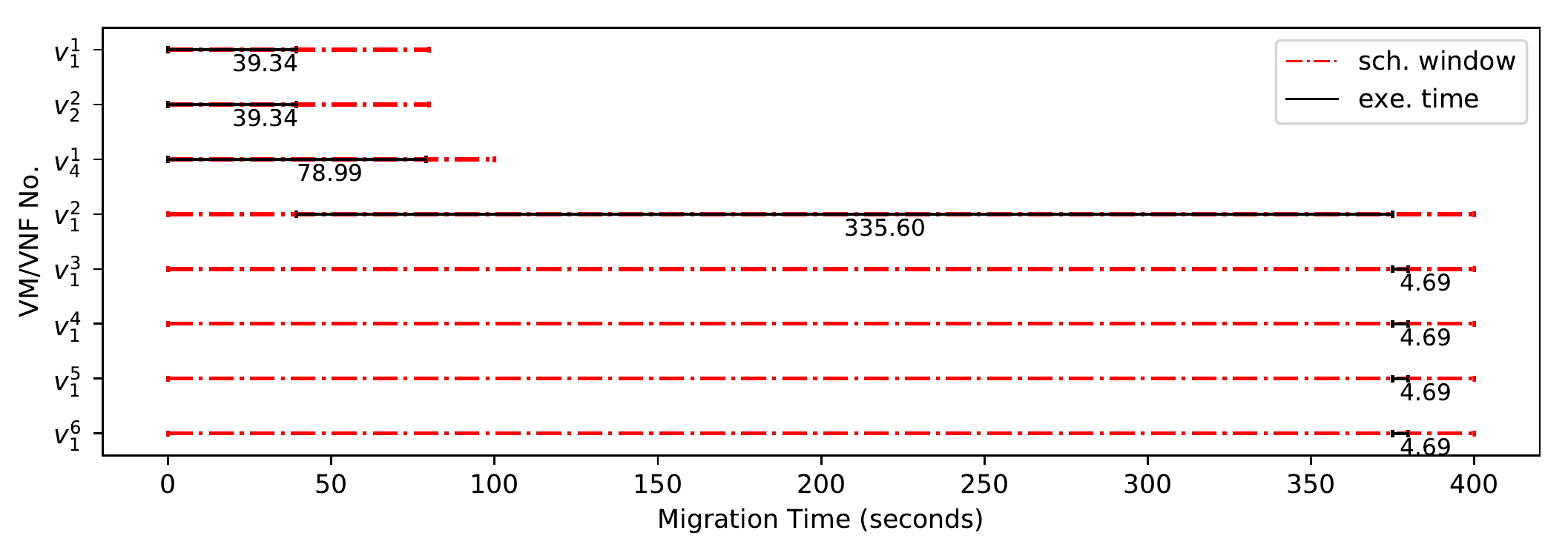}
		\caption{migration order $S_1$}\label{fig: case-study-o1}
	\end{subfigure}
	\begin{subfigure}[t]{.9\textwidth}
		\centering
		\includegraphics[width=\textwidth]{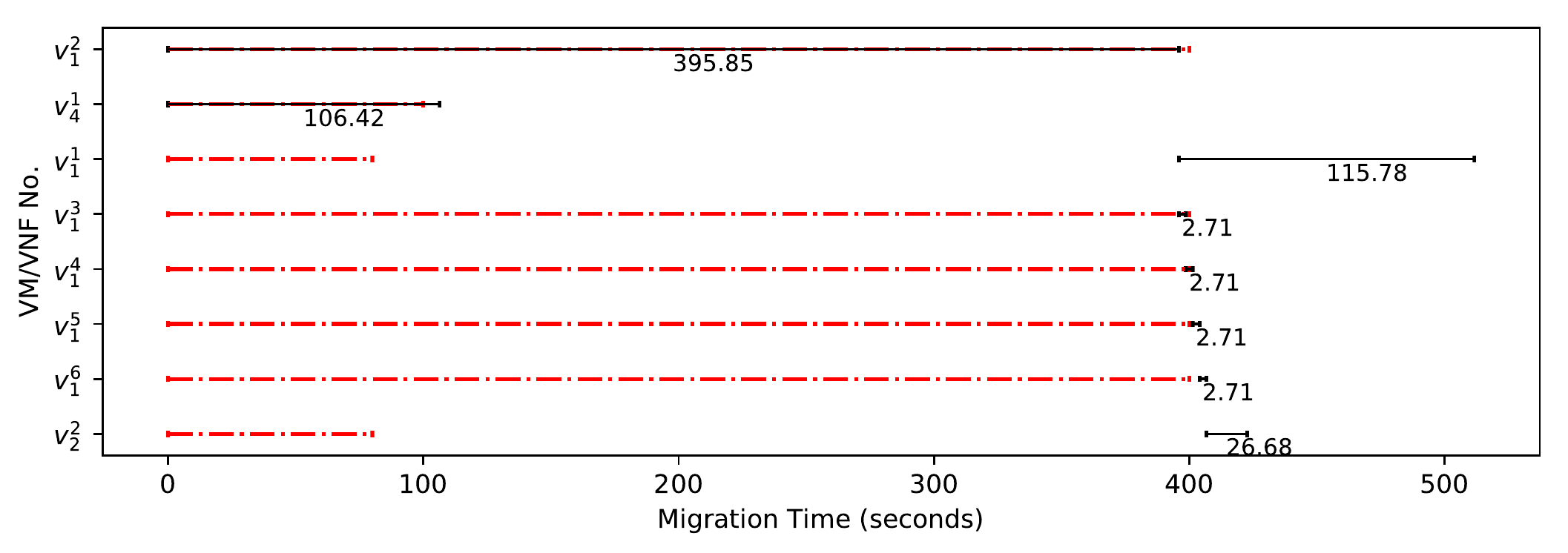}
		\caption{migration order $S_2$}\label{fig: case-study-o2}
	\end{subfigure}	
	\caption{Results of different scheduling orders} \label{fig: case-study-orders}
\end{figure}

At the time $t_0=0$, the coordinator receives several migration requests at the same time based on the configured optimal reallocation interval as shown in Fig. \ref{fig: case-study}. Other VDCs and SFCs which are unrelated to the migration in the host are not shown. The maximum memory copy round is 30 and the downtime threshold is 0.5 second. There are two of possible orders: ${S_1} = (v_1^1,v_2^2,v_4^1),(v_1^2),(v_1^3,v_1^4,v_1^5,v_1^6)$ and ${S_2} = (v_1^2,v_4^1),(v_1^1,v_1^3)$,\\ $(v_1^4),(v_1^5),(v_1^6),(v_2^2)$, shown in Fig. \ref{fig: case-study-orders}.
Migration tasks within the same group could perform concurrently. For subsequent migrations from different concurrent migration group, the scheduler will start a migration as soon as a sharing-resource migration in the other group is finished. For example, migration of $v_1^2$ will start after the migration of instance $v_1^1$ is finished (Fig. \ref{fig: case-study-o1}). After each migration, all virtual links connected to the migrated instance will be rerouted to the destination host. 
Therefore, the available bandwidth of the remaining migrations will be updated according to the new instance placement at the end of each migration. 


By leveraging simulation capabilities for both computing and networking, we implemented and extended the corresponding components based on the CloudSimSDN \cite{cloudsimsdn-nfv} to simulate each phase of pre-copy live migration.
As shown in Fig. \ref{fig: case-study-orders}, the total migration time $T_{total}$ of two migration scheduling orders is 377.645 and 511.625 seconds, respectively. The average downtime $\Sigma T_d/n$ is 0.317 and 0.353 with maximum 0.807 and 1.538 seconds for $v_1^2$ instance. Furthermore, for the migrations of $v_1^3,v_1^4,v_1^5,$ and $v_1^6$ instance, as the processing time of computing overheads is 2.0 seconds, the total migration time from the start of migration $v_1^3$ is 4.691 and 10.593 seconds by using parallel and sequential method, respectively. The average value of the remaining scheduling window $\Sigma \xi/n$ of two orders is 26.104 and -99.708, respectively. Although these orders both perform concurrent migrations that do not share the same resources, the first scheduling order leads to a better performance in terms of total migration time, average downtime, SFC/VFC migration time, and remaining scheduling window (i.e. less SLO violations).

\subsection{Impact of Bandwidth and Dirty Rate}\label{sec: impact-bw-dirty-rate}
\begin{figure}[t]
	\centering
	\includegraphics[width=0.4\linewidth]{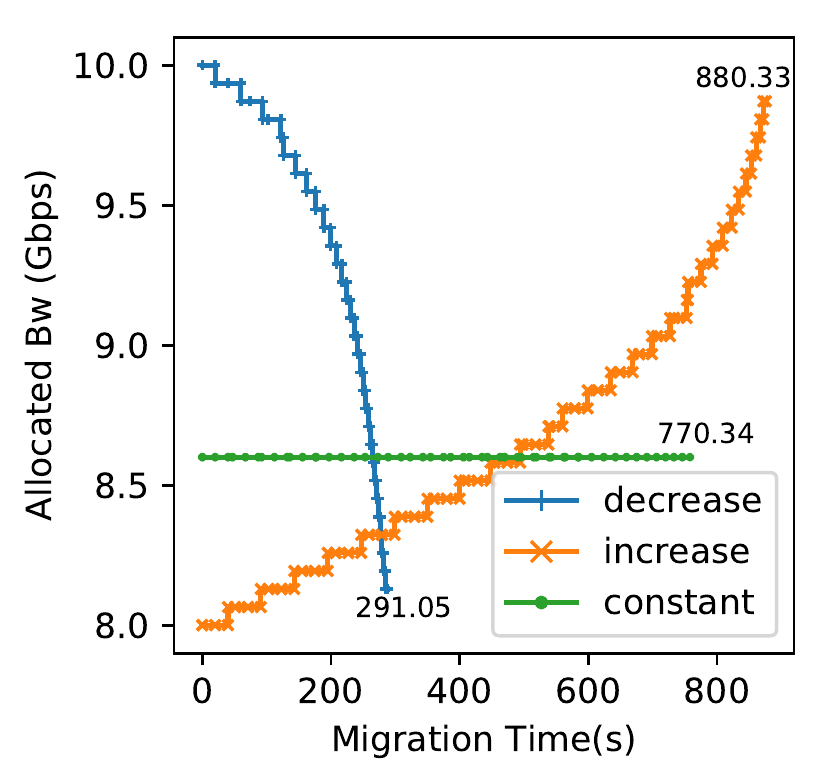}
	\caption{Migrations with xlarge flavor under various bandwidth functions}
	\label{fig: bw-steps}
\end{figure}

First, we argue that the bandwidth allocated to the early iterative transmission rounds can highly affect the performance of a migration. Based on the mathematical model shown in Section \ref{sec:single-model}, Figure \ref{fig: bw-steps} illustrates the migration performance under three different bandwidth functions, where: 1) begins with low bandwidth then increases the bandwidth for each iteration round; 2) has a constant bandwidth; 3) starts with high bandwidth then decreases the bandwidth for each iteration round. It indicates that even with the same average bandwidth during the migration, insufficient bandwidth in the early steps will cause a huge amount of dirty pages remained to transfer for the subsequent transmission rounds (Equation (\ref{eq: var-mem-copy})). This causes a much slower convergence process to reach the point that remaining dirty pages satisfies the downtime threshold. Furthermore, according to the migration threshold and round constraints (Equation (\ref{eq:round-number})), starting the migration with insufficient bandwidth results in a large accumulation of the remaining dirty pages in the previous rounds. In other words, in order to complete the migration within a reasonable time, an unacceptable service downtime will occur regardless of the downtime threshold due to the migration round constraint. 

\begin{table}[t]
	\caption{List of commonly used notations}
	\centering
	\label{tbl:notations}
	\resizebox{0.5\linewidth}{!}{
	\begin{tabular}{|l||l|}
		\hline
		Notation	&	Description\\
		\hline
		M          & Memory size of VM/VNF          \\
		R          & Dirty page rate                \\
		L          & Bandwidth assigned to Migration \\
		$T_{pre}$ & Pre-migration processing time  \\
		$T_{post}$ & Post-migration processing time \\
		$T_{network}$ & memory copy network transmission time \\
		$V_i$	& the transferred data of $i_{th}$ round of memory copy \\
		$T_i$	& the time interval of  $i_{th}$ round of memory copy \\
		$T_{dthd}$ & the downtime threshold \\
		$\rho $		& the memory compression rate \\
		$\sigma$	& the ratio of R to L multiple $\rho$ \\
		$\lambda(p)$		& the maximum allowed parallel number in path $p$ \\
		$r$		& the processing speed of one compute core \\
		$T_{network}^n$ & parallel $T_{network}$ of n migrations in the same path \\
		$m_j$	& the memory size of migration j; \\
		$s_j, d_j$	& the ordered pair of source and destination\\
		$N$	& the set of physical network nodes \\
		$E$	& the set of physical network links \\
		$N^i$	& nodes set of VDC/SFC $G^i$ \\
		$E^i$	& links set of VDC/SFC $G^i$ \\
		$P$ 	& set of all paths in the network \\
		$P_j$	& set of all paths between $(s_j, d_j)$	\\ 
		$c(e)$	& capacity of link e	\\
		$u(e)$	& residual bandwidth in link $e$ \\
		$u(p)$	& available bandwidth of path $p$ \\
		$D_j$	& deadline of migration $j$ \\
		$D(G^j)$	& deadline of all migrations that $n_j \in G^i$ \\
		$\theta$	& maximum tolerant number of SLA violations \\
		$\omega$	& cumulative SLA violation rate \\
		$\xi$	& remaining migration scheduling window\\
		$Y_t$	& cumulated violations at time stamp $t$ \\
		\cline{1-2}
	\end{tabular}
}
\end{table}

\subsection{Deadline-related Migration}
In this section, we discuss deadline-related migration. 
As one of the reasons for SLA violation, the Service Level Objective (SLO) violation speed and total violation threshold are the main monitoring parameters. If the cumulative violations of SLO  exceed the threshold, one SLA violation happens. In this situation, the migration request is generated due to the SLO violations under the current placement of VMs/VNFs with the workload bursting, end-user mobility, and resource over-subscription \cite{cziva2018dynamic, son2017sla, guo2014cost}. The SLO thresholds are configured by the cloud infrastructure management to alert on the significant events. A migration request can be issued due to the SLO violations under the current situation. For instance, VNF need to be relocated to the cloud before exceeding the threshold of cumulative violation due to current SLO violation rate in terms of response time and latency \cite{cziva2018dynamic}, and VMs need to be migrated from the overbooked hosts due to the service workload burst \cite{guo2014cost}. The burst of workload and location changing of end users can cause serious SLA violations and QoS degradation. Thus, VM/VNF migrations need to be finished before a certain deadline to prevent the cumulative SLO violations exceeding the threshold.

Thus, the deadline for the VM migration can be estimated based on the threshold of total allowed SLO violations and the current SLO violation speed. Based on the new optimal allocation, the placement algorithm will request corresponding VM migrations to prevent the accumulated SLO violations from exceeding the threshold. 
Among these migration tasks, different services and VNFs have several levels of urgency in terms of the current average SLO violation cumulative rate $\omega$ based on the monitoring, the current number of cumulated violations $Y_t$, and the threshold of total violations $\theta$. Therefore, when the dynamic resource policy triggered  at time $t$ by the configured period, the deadline of migration task can be calculated as:
\begin{equation}\label{eq: slo-deadline}
D = {{\left( {\theta  - {Y_t}} \right)} \mathord{\left/
		{\vphantom {{\left( {\theta  - {Y_t}} \right)} \omega }} \right.
		\kern-\nulldelimiterspace} \omega }
\end{equation}
For the migration tasks which specify the scheduling window (e.g., scheduled maintenance), the deadlines can be directly used as the input for migration scheduling. 

Furthermore, there are time-critical migration requests with specific deadline $D(G^i)$ for the whole SFC or VDC $G^i$. In other words, all related VMs/VNFs inside the SFC/VDC need to be migrated and run in the destination hosts before the required deadline. A simple solution is to directly assign the group deadline to each migration task. For better performance, the deadline for each task can be calculated by subtracting the sum of the worst execution time of other migration tasks from the group deadline: 
\begin{equation}\label{eq: group-deadline}
{D_k} = D\left( {{G^i}} \right) - \sum\nolimits_{{n_j} \in {{{G^i}} \mathord{\left/
			{\vphantom {{{G^i}} {{n_k}}}} \right.
			\kern-\nulldelimiterspace} {{n_k}}}} {T_{exe}^j} 
\end{equation}

In practice, live VM migrations can be scheduled in low activity periods \cite{tsakalozos2017live}. VMs or VNFs interact with different groups of end users with geographical variances or applications with different access features. For instance, VMs of Web applications allocated in the same physical host serve different areas, such as China, Japan, Australia, and Europe, may experience hours or minutes low-activity scheduling window. As a large amount of VMs/VNFs with various features allocated in relative limited physical hosts, the low activity window for migration scheduling can be extremely limited. 

\subsection{Problem Formulation}
In this section, we formally define the problem of live SFC/VDC migration planning as a Mixed Integer Linear Programming (MILP) problem. In the model, the physical data center is represented by a graph $G=(N,E)$, where $N$ denotes the set of nodes including physical hosts, switches, and routers, and $E$ denotes the set of directed links between nodes. The remaining CPU, memory, and disk in the destination node $N$ should be larger than the resources required by the migrating VM. 

Let $\tau$ denote the instant of time when a migration starts or finishes. From the beginning of the first migration to the end of the last migration, at least one migration is is in progress in the data centers. Let $T_{mig}^{i}$ denote the response time (single execution time of the migration $i$ plus the waiting time to start). Then, for more concise expression, we use $\tau_0$ as the begin time instant and $\tau_K$ as the end time instant for a total of $K$ migration tasks.

\begin{equation}\label{eq: tau}
{\tau _i} \in \left[ {0,T_{mig}^1,...,T_{mig}^K} \right] = \left[ {{\tau _0},{\tau _1},...,{\tau _K}} \right]
\end{equation}
where the total $K$ migrations are sorted by the completion time and ${\tau _i} \ge {\tau _{i - 1}},i = 0,1,...,K$. It converted the original problem to total K discrete state.

Let $X_k^{{\tau _i}} \in \left\{ {0,1} \right\}$ denote the binary variable that indicates whether the migration $k \in {\mathbb{R}^ + }$ occurs at time interval $\tau \in \left[ {{\tau _i},{\tau _{i + 1}}} \right)$. Therefore, the response time of migration k can be calculated as:
\begin{equation}\label{eq: tau-k}
{\tau _k} = T_{mig}^k = \sum\limits_{i = 1}^k {X_k^{{\tau _i}} \cdot \left( {{\tau _i} - {\tau _{i - 1}}} \right)} ,1 \le k \le K
\end{equation}

As mentioned in section \ref{sec:single-model}, the migration task cannot be preempted (stopped) after it is started. For memory state synchronization, the transferred memory data (dirty pages) in previous iterative rounds will become infeasible and cause an unacceptable overhead on the DCNs.
Thus, we need to add the following constraint to the binary variable:
\begin{equation}
X_k^{{\tau _i}} \ge X_k^{{\tau _{i - 1}}},0 \le {\tau _i} < {\tau _k}
\end{equation}
Furthermore, let ${Z_{j,k}}$ denote the binary variable indicating whether two migrations $j$ and $k$ can be performed concurrently:
\begin{equation}
{Z_{j,k}} = {X_j} \cdot {X_k} = \left\{ {\begin{array}{*{20}{c}}
	1,\\
	0,
	\end{array}} \right.\begin{array}{*{20}{c}}
{indep.}\\
{sharing.}
\end{array}
\end{equation}
If migration $j$ and $k$ share the same pair of source and destination or network paths, thereby affecting the available bandwidth allocated to either migration, the two migrations are resource dependent (sharing). Otherwise, two independent migrations can be performed concurrently.

Let $P_{k}$ denote the set of paths $p$ available for the migration $k$. The relation between allocated bandwidth for migration $k$ and available bandwidth along the path $p$ can be represented as:
\begin{equation} \label{eq: l-mig}
{l_k} = \sum\nolimits_{p \in {P_k}} {x\left( p \right)} 
\end{equation}

 According to the SFC/VDC $G^j$ and physical DCN $G$, the total flows including migration transmission $p$ and reserved service virtual links $p'$ can not exceed the capacity $u(e)$ of link $e$. For $\forall {\tau _i},i = 0,1,...,K$, we calculate the available bandwidth for migration $l_k^{\tau_i}$ under the new input because the migration $i$ is finished at time instant $\tau_i$ and the virtual links need to be rerouted due to the new placement. The constraint during time interval $\tau = \left[ {{\tau _i},{\tau _{i + 1}}} \right)$ can be represented as follows:
\begin{equation} \label{eq: l-mig-sfc}
\sum\nolimits_{p \in {P_e}} {x\left( p \right)}  + \sum\nolimits_{p' \in {P_e}} {x\left( {p'} \right)}  \le u\left( e \right),\forall e \in E
\end{equation}

Moreover, the allocated bandwidth for migration $k$ cannot exceed the interface capacity of source and destination hosts. There is no allocated bandwidth before the migration begins and after it is completed. Thus, we have the constraints expression as follows:
\begin{equation} \label{eq: l-hostc}
{l_k} \le \min \left\{ {C_s^k,C_d^k} \right\}
\end{equation}
\begin{equation} \label{eq: l-maxc}
l_k^\tau  \le X_k^\tau  \cdot \Psi 
\end{equation}
where $C_s^k,C_d^k$ denote the interface capacity of source and destination host. $\Psi  \in {\mathbb{R}^ + }$ is a constant larger than the maximum bandwidth of paths in the network that could be allocated to the migration. 

In addition, as shown in Section \ref{sec: impact-bw-dirty-rate}, if the allocated bandwidth for the first few transmission rounds is smaller than the dirty page rate, there will be a huge performance degradation. Thus, we add the extra constraint to $l_k^{\tau_0}$ for the migration start:
\begin{equation}\label{eq: tau-r}
l_k^{{\tau _0}} > {r_k}
\end{equation}

The problem of minimizing the total migration time and SLO violations during the scheduling can be formulated as:
\begin{equation}
\min \left( {\sum\limits_{i = 1}^K {X_K^{{\tau _i}}}  \cdot \left( {{\tau _i} - {\tau _{i - 1}}} \right) + \sum\limits_{k = 1}^K {\left( {{\tau _k} - {D_k}} \right)} } \right)
\end{equation}
subject to constraints (\ref{eq: tau})-(\ref{eq: tau-r}). 
The commonly used notations in the paper are shown in Table \ref{tbl:notations}.

The problem is NP-hard to solve because it generalizes the data migration problem \cite{bari2014cqncr} without the extra constraints of resources and migration deadline. The model in \cite{wang2017virtual} also represents the same problem, but it didn't consider the impact of flow relocation on the performance of remaining migrations. They are all proved to be NP-hard. Solving the MILP problem in a reasonable time is not feasible, because the general algorithms supported in MILP solver will lead to extremely high time complexity.

\section{Algorithm Design}
In this section, we describe the details of our algorithm. The proposed deadline-aware multiple migration planning algorithm has two main components: the migration group planning and the on-line scheduler.
Observations and algorithm rationales are as follows:
\begin{itemize}
	\item Since live migration is highly dependent on available network bandwidth, migrations with different network paths, source and destination hosts can be performed concurrently. The scheduling algorithm should maximize the number of resource-independent tasks migrating at the same time. In addition, for a single migration, multi-path transmission can improve performance.
	\item Due to the computational overhead, migrations with low dirty page rate and small VM memory size can be migrated in parallel through the same paths by treating them as one migration \cite{he2019performance}. On the other hand, for migrations with large memory and dirty page rate, the sequential schedule for resource-dependent migrations can optimize the total migration time.
	\item One physical host interface can only receive one and send one migration at the same time, i.e, one pair of ordered source and destination hosts $(s_j,d_j)$ can only be assigned to one migration at the same time.
	\item After each migration completion, the network resources used by both migrations and cloud services will change. For migration plans such as consolidation, migrations with small execution time quickly free up more bandwidth for subsequent migrations, thereby reducing the total migration time. On the other hand, migrations that negatively affect network bandwidth will increase the execution time of other migrations.
	\item If the available bandwidth is smaller than the dirty page rate, the migration should not be started as the accumulated dirty pages will become bigger after each round of memory copy, resulting in unacceptable migration execution time and downtime.

\end{itemize}

\subsection{Multiple Migration Planning}

The proposed heuristic algorithm for concurrent migration group is shown in Algorithm \ref{alg: main-alg}. Given the input of migration requests in terms of flavor, dirty rate, compression ratios, the scheduling windows of migration tasks (deadlines), and the pair of source and destination host, the algorithm will return the ordered list of concurrent migration groups where each group is the maximal resource-independent migration group.

\begin{algorithm}[htbp]
	\caption{Heuristic graph-based algorithm of concurrent migration grouping}\label{alg: main-alg}
	\SetKwFunction{conG}{\textsc{Get}ConcurrentGroup}
	\SetKwFunction{isIndep}{\textsc{Is}Independent}
	\SetKwFunction{completeG}{\textsc{Create}CompleteDepGroup}
	\KwIn{\{$n:s_n \rightarrow d_n$\}}
	\KwResult{migGroups $\{G_{mig}^{S_q}\}$}
	\{Creating Dependency Graph $G_{dep}$ of Mig Tasks\}\\
	$G_{dep} \gets null$;\\
	\ForEach{$n_j$ in FeasibleMigs}{
		\For{$n_{j+1}$ in FeasibleMigs}{
			\If{\nonl\isIndep{$m_k,m_j$}==$0$}{
				addEdge($G_{dep}$,($m_k,m_j$));\\
			}
		}
	}
	\{Creating resource-dependent complete subgraphs\}\\
	$\{N_{dep}\} \gets \emptyset$;\\
	\ForEach{$n_j \in G_{dep}$}{
		\If{IsVisited($n_j$)==$False$}{
			$N_{dep}^j \gets \{n_j\}$;\space\space\space\space{//}complete graph contains mig $j$;\\
			SetVisited($n_j$) $\gets Ture$;\\
			\nonl\completeG{$n_j,G_{dep},N_{dep}^j$};\\
			$\{N_{dep}\} \gets \{N_{dep}\} \cup N_{dep}^j $;\\
		}
	}
	
	\{Scoring and Sorting each node\}\\
	\ForEach{$N_{dep}^i \in \{N_{dep}\}$}{
		\ForEach{$n_j$ in $N_{dep}^i$}{
			$cost\left( {{n_j}} \right) \leftarrow \alpha  \cdot T_{mig}^j + \beta  \cdot \left( {T_{mig}^j - {D_j}} \right) + \gamma  \cdot {I_j}$;\\
		}
		$N_{dep}^i \gets$ sorting($N_{dep}^i,$\{cost($n_j$)\});\\
	}
	\{Get migration groups from node-weighted subgraphs\}\\
	\KwRet{$\{G_{mig}^{S_q}\}$} $\gets$ \nonl\conG{$\{N_{dep}\}$};\\
\end{algorithm}

\begin{algorithm}[htbp]
	\caption{Creating concurrent migration groups}\label{alg: main-alg2}
	\SetKwFunction{conG}{\textsc{Get}ConcurrentGroup}
	\nonl\conG{$\{N_{dep}\}$}:\\
	$S_q \gets 0$;\space\space\space\space{//}scheduling priority for migration groups;\\
	$G_{mig}^{S_q} \gets \emptyset$;\\
	\ForEach{$n_j \in \{N_{dep}\}$}{
		$new \gets Ture$;\\
		\For{$s=0$ to $S_q$}{
			$flag \gets True$;\\
			\ForEach{$n_k \in G_{mig}^s$}{
				\If{getEdge($n_j,n_k,G_{dep}$)}{
					$flag \gets False$;\\
				}
			}
			\If{$flag == True$}{
				$G_{mig}^s \gets G_{mig}^s \cup \{n_j\}$\\
				$new \gets False$;\\
				\Break;\\
			}
		}
		\If{$new == True$}{
			$S_q \gets S_q+1$;\\
			$G_{mig}^{S_q} \gets \{n_j\}$;\\
		}
		delete($G_{dep},\{N_{dep}\},n_j$);\\
	}
	sorting($\{G_{mig}^{S_q}\}, \sum cost(n_j \in G_{mig}^s)$);\\
	\KwRet{$\{G_{mig}^{S_q}\}$}\\
\end{algorithm}

First, we need to assign the deadline to each SLO-related migration task based on the Equation (\ref{eq: slo-deadline}) and (\ref{eq: group-deadline}).
Secondly, considering the computing overheads, the migration tasks need to preprocess the integrated network-sharing migrations that suit the parallel method. In other words, if the migration time not exceeds the deadline and the total migration time is reduced, the scheduler will perform the parallel method for such migration.

From \textbf{line 2-5} in Algorithm \ref{alg: main-alg}, the dependency graph $G_{dep}$ is created for all feasible migration tasks. If two tasks share migration resources (dependent), the edge $(n_j,n_k)$ will be added into $G_{dep}$. As we allow multi-path transmission for memory copying, not only the ordered pair of source and destination $(s,d)$ but also intersected network paths of migrations with different $(s,d)$ will be shared. Therefore, whether two migrations can be concurrently performed is described in Algorithm \ref{alg:sharing}, where $P_k$ denotes the set of paths that can be allocated to migration $k$, $u(P_k)$ denotes the total available bandwidth, and $C_s^k,C_d^k$ denote the interface capacity of source and destination hosts. In addition, $X_j*X_k=0$ denotes that migration $j$ and $k$ share resources (dependent). Otherwise, the two migrations with $X_j*X_k=1$ can be performed concurrently.

\LinesNumberedHidden
\begin{algorithm}[htbp]
	\SetAlgoLined
	\DontPrintSemicolon
	
	\KwIn{$P_k,P_j,(s_k,d_k),(s_j,d_j)$}
	\KwResult{$X_j*X_k=1$ or $0$}
	
	\SetKwFunction{FMain}{IsIndependent}
	\SetKwProg{Fn}{Function}{:}{}
	\Fn{\FMain{$m_k,m_j$}}
	
	\eIf{${s_k} == {s_j}$ \textbf{and} ${d_k} == {d_j}$}{
		\KwRet $X_j*X_k=0$;
	}{
	\eIf{${s_k} \ne {s_j}$ \textbf{and} ${d_k} \ne {d_j}$}{
		\If{$u\left( {{P_j}} \right) - u\left( {{P_j} \cap {P_k}} \right) \ge \min \left( {u\left( {{P_j}} \right),C_s^j,C_d^j} \right)$ \textbf{and} $u\left( {{P_k}} \right) - u\left( {{P_k} \cap {P_j}} \right) \ge \min \left( {u\left( {{P_k}} \right),C_s^k,C_d^k} \right)$}{
			\KwRet $X_j*X_k=1$;	
		}
	}
	{
		\KwRet $X_j*X_k=0$;
	}}
	
	\caption{Check independence of two migrations with multi-paths and interface constraints}\label{alg:sharing}
\end{algorithm}

From \textbf{line 7-13} in Algorithm \ref{alg: main-alg}, we divide the dependency graph $G_{dep}$ into the largest complete dependency subgraph of the remaining graphs $\{N_{dep}\}$, where each migration is dependent on others. One migration $j$ exist and can only exist in one complete subgraph $n_j \in N_{dep}^i$ as the complete dependency subgraph $|N_{dep}|$ is the largest. Between complete subgraphs, there are links remained according to the original dependency graph. The corresponding recursive algorithm is described in Algorithm \ref{alg:dep-graph}.
\begin{algorithm}[ht]		
	\caption{Create Complete dependency Subgraph}\label{alg:dep-graph}
	\SetKwFunction{completeG}{\textsc{Create}CompleteDepGroup}
	\nonl\completeG{$n_j,G_{dep},N_{dep}^j$}:\\
	\Indp
	$N_{adj}^j \gets$ adjacency($G_{dep},n_j$);\\
	SetVisited($n_j$) $\gets Ture$;\\
	\For{$n_k \in N_{adj}^j$}{
		\If{IsVisited($n_k$)==$False$ \textbf{and} IsCompleteGraph($N_{dep}^j,n_j$)}{
			$N_{dep}^j \gets N_{dep}^j \cup \{n_k\}$;\\
			\completeG{$n_k,G_{dep},N_{dep}^j$};\\
		}
	}
	\If{$|N_{dep}^j|$ larger than previous}{
		\KwRet{$N_{dep}^j$}
	}
\end{algorithm}

From \textbf{line 15-18}, in each complete subgraph, we calculate the score of each migration (line 17) and sort them from the smallest to the largest based on the score. For the function of migration cost $cost(n_j)$, it is the weighted sum of the migration time (Equation (\ref{eq:total-com-mem}), (\ref{eq:mem-copy}), and (\ref{eq:round-number})), minus slack time, and the impact of migration $j$ on other migrations, where $\alpha$, $\beta$, $\gamma$ are coefficients. 
In our algorithm, as the cost of each individual migration is evaluated separately, we categorize the benefit of single migration into direct and potential impact ${I_j} = a \cdot I_j^{direct} + b \cdot I_j^{potent}$, where $a+b=1$. The direct impact of migration $j$ can be represented as:
\begin{equation}
\begin{array}{l}
{I^{direct}_j} = \left( {\sum\limits_{{n_k} \in \left\{ {{N_{dep}}} \right\} - {n_j}} {T{{_{exe}^k}^\prime } - \sum\limits_{{n_k} \in \left\{ {{N_{dep}}} \right\}} {T_{exe}^k} } } \right)\\
+ \left( {\sum\limits_{{n_k} \in \left\{ {{N_{dep}}} \right\} - {n_j}} {\left( {T{{_{mig}^k}^\prime } - {D_k}} \right) - \sum\limits_{{n_k} \in \left\{ {{N_{dep}}} \right\}} {\left( {T_{mig}^k - {D_k}} \right)} } } \right)
\end{array}
\end{equation}
where
$\{N_{dep}\}$ is the set of all complete dependency subgraphs. $T{{_{exe}^k}^\prime}$ and $T{{_{mig}^k}^\prime}$ denotes the execution time and the migration time after the migration $n_j$ is completed. If the migration $n_k$ and $n_j$ are resource dependent, $T{{_{mig}^k}^\prime}$ will be the sum of $T{{_{exe}^k}^\prime}$ and $T_{exe}^j$.

The potential impact considers the possibility of decreased migration time when the bandwidth of some parts of the migration paths increases. Then, it can be represented as:
\begin{equation}
I^{potent}_j = \sum\limits_{{n_k} \in \left\{ {{N_{dep}}} \right\} - {n_j}} {\sum\limits_{p \in {P_k}} {\frac{{\left| {\{ \hat e\} } \right|}}{{\left| p \right|}} \cdot \left( {T_{mig}^{k,u(\bar e)} - T_{mig}^k} \right)} }
\end{equation}
where ${\left| {\{ \hat e\} } \right|}$ is the number of links with increased bandwidth and $\hat e \in p$, $p \in {P_k}$. The migration time ${T_{mig}^{k,u(\bar e)}}$ is based on the minimal increased bandwidth among the links $u(\bar e) = \min (u(\hat e))$.


In the final step (line 18), the cost-driven algorithm creates concurrent migration groups (Algorithm \ref{alg: main-alg2}), where the selected migrations are resource independent. As shown in Algorithm \ref{alg: main-alg2}, according to the sorted $N_{dep}^j \in \{N_{dep}\}$, it will always first select a migration $n_j$ with the lowest score from each complete dependency subgraph $N_{dep}^j \in \{N_{dep}\}$ (line 3). If there is no migration group feasible for $n_j$ ($new == true$), it will create a new concurrent migration group $G_{dep}^s$. After adding the migration to one migration group $G_{mig}^s$, it will be deleted from the dependency graph $G_{dep}$ and the corresponding subgraph $N_{dep}^j \in {N_{dep}}$ (line 17). In line 18, migration groups are added to the sorted list from minimum to maximum score in seconds.

When additional migration tasks arrive after the initial processing, the on-line migration scheduler will first remove the node from the migration dependency graph after completing one migration.
If additional migration tasks arrive, our proposed algorithm will add the new tasks to the existing migration dependency graph. The planning algorithm will also remove the ongoing migrations from the dependency graph. Then, it recalculates the plan based on the current system status (available network and computing resources).

\subsection{Time Complexity Analysis}
Let $N$ denote the total migration tasks number. Then, the process for creating dependency graph requires $O\left( N \right)$. For the breadth-first research in dependence graph to create complete subgraphs, it requires $O\left( {N + {{N\left( {N - 1} \right)} \mathord{\left/
			{\vphantom {{N\left( {N - 1} \right)} 2}} \right. 
			\kern-\nulldelimiterspace} 2}} \right)$.
Let	$E$ denote the total number of physical links. Then, the time complexity of cost function (Line 16) is $O\left( {NE} \right)$.
Thus, The worst case time complexity of scoring and sorting is
$O\left( {{N^2}E + N\log \left( N \right)} \right)$.
The worst case for creating concurrent migration group is $O\left( {{N^2}} \right)$. Therefore, the time complexity of worst case of Algorithm \ref{alg: main-alg} is $O\left( {{N^2}E} \right)$.

\subsection{On-line Migration Scheduler}
\begin{algorithm}[htbp]
	\caption{Updating and scheduling feasible migrations}
	\label{alg:scheduler-1}
	\KwData{migGroups, currentGroupNum}
	\KwResult{Start feasible migrations and groups}
	\ForEach{G in migGroups}{
		groupNum = getGroupNum(G);\\
		\If{groupNum $\le$ currentGroupNum}{
			\ForEach{mig in G}{
				\If{isMigFeasible(mig)}{
					processMigrationStart(mig);\\
				}
			}
		}
	}
	\{Scheduling migration in subsequent group\};\\
	\eIf{hasNext($G_{current}$)}{
		$G_{next}$ = getNextGroup($G_{current}$);\\
		flag = $False$;
		\For{mig in $G_{next}$}{
			\If{isMigFeasible(mig)}{
				preocessMigrationStart(mig);\\
				flag = true;\\
			}
		}
		\If{flag}{
			$G_{current}$ = $G_{next}$;
		}
	}
	{
		\If{size(inMigrationList) == 0 and size(migPlan)==0}{
			setTotalMigTime(migPlan);\\
		}
	}
	
\end{algorithm}	
For the real data center environment, the network workloads vary greatly over time. Therefore, it is impracticable to set the start time of each migration just based on the prediction model and the available bandwidth at the current time $\tau=0$. The proposed on-line migration scheduler can dynamically schedule the subsequent migrations at the end of each migration. 	

The algorithm used by the SDN-enabled migration scheduler is shown in Algorithm \ref{alg:scheduler-1}. It includes two steps: 1) check feasible migrations in the previous and current migration groups; 2) start all feasible migrations in the next group. By only considering to start the next migration group in the ordered list at each time, it prevents the occurrence of priority inversion. The priority inversion refers to the migration group with a higher score (lower priority) may start to migrate before the group with a smaller score.

\section{Performance Evaluation}
In this section, we first introduce the configuration of our event-driven simulation system, then the various scenarios and parameters to be evaluated in both inter and intra-datacenter environments. In the end, we analyze the results and conclude the experiments. We compare the performance of SLAMIG with the one-by-one scheduling and other two state-of-art algorithms \cite{bari2014cqncr, wang2017virtual}.  The results indicate that our proposed algorithms achieve good migration performance in terms of the total migration time, total transferred data, average migration time, and average downtime, meanwhile can efficiently reduce the deadline violations during the multiple live migrations. Furthermore, we evaluate and analyze the impact of multiple migration planning and scheduling on the energy consumptions and the QoS of applications.

\subsection{Simulation System Configuration}
\begin{figure}[th]
	\centering
	\includegraphics[width=0.5\linewidth]{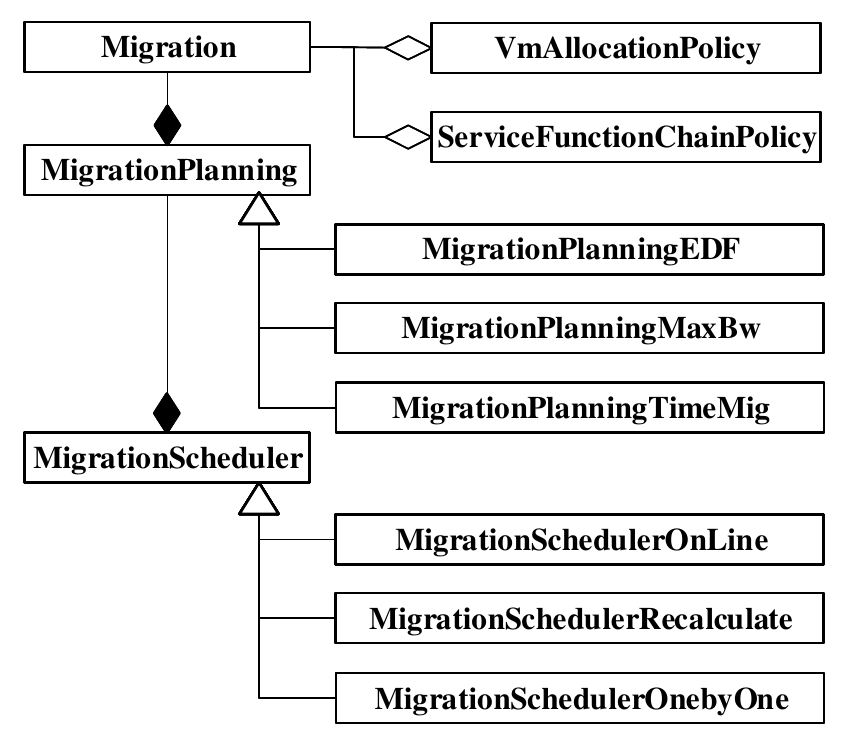}
	\caption{MigrationScheduler to simulate single live migration follows the sequence of MigrationPlanning}
	\label{fig: uml-mig}
\end{figure}

\begin{table}[th]
	\caption{Simulation events in \textit{MigrationScheduler}, \textit{SDNDataCenter}, and \textit{NetworkOperatingSystem}}
	\resizebox{\linewidth}{!}{
		\begin{tabular}{|c|l|c|}
			\hline
			Num. &Event and Operation	&	Function	\\
			\hline
			0&$SDN\_VM\_MIG\_PRE$ & check available network and set up the migration routing \\
			1&$SDN\_VM\_MIG\_START$ & start the pre-copy phases \\
			2&$SDN\_PACKET\_COMPLETE$ & check the application and migration flows, estimate the downtime and send the remaining
			dirty page\\
			3&$SDN\_PACKET\_SUBFLOW\_COMPLETE$ & check the completion of multiple migration flows\\
			4&$SDN\_VM\_PAUSE$ & pause the VM/VNF based on the downtime and iteration threshold\\
			5&$SDN\_VM\_RESUME$ & resume the VM/VNF on the dest host after the completion of the stop-and-copy flow \\
			6&$SDN\_VM\_MIG\_POST$ & shut and delete the original instance and rerouting the flows to the new VM/VNF.\\
			7&$SDN\_VM\_MIG\_SCHEDULER$ & process the migration scheduling in the current time.\\
			\hline
		\end{tabular}
	\label{tb: sim-events}
	}
\end{table}

\begin{table}[th]
	\caption{Parameters supported in event-driven simulator}
	\resizebox{\linewidth}{!}{
		\begin{tabular}{|l||l|l|l|l|l|l|l|}
			\hline
			Type 	   &  	\multicolumn{7}{c|}{Parameters}   \\
			\hline
			\textbf{computing}  & CPU  & Memory 	 & Disk	&  Workloads	 & 	Task Scheduling	& Task Priority	    & Overbooking Ratio  \\
			\textbf{networking} & Bandwidth	  & Topology	 & 	Switch Buffer & Ports & Channel & Control Plane	   &  Data Plane 	  \\
			\textbf{monitoring} & Statistic  & Energy Consumption & 	Utilization &  Response time & Network Delay & Fault Handling & 	 \\
			\textbf{live migration}  & dirty page rate	&  mig. time  & downtime 	 & 	transferred data   & deadline &  available bw	&  flow path	  \\
			\hline
		\end{tabular}	
	}
	\label{tb: parameters}
\end{table}

In this section, we first introduce the details of our experiment platform for SDN and NFV-enabled cloud data centers.
To evaluate the performance of large-scale multiple live migrations, we extended the CloudSimSDN-NFV \cite{cloudsimsdn-nfv} by implementing the phases of live VM migration and corresponding parameters (Table \ref{tb: sim-events} and Fig. \ref{fig: migration}). It is an event-driven simulation environment supporting SDN-enabled cloud computing. It also provides the mechanism of auto-scaling of VNF and automatic load balancing through different SFCs. Table \ref{tb: parameters} illustrates some parameters supported by the extended version. 

Fig. \ref{fig: uml-mig} illustrates the implemented components regarding live VM migration: 
\textit{Migration} Class contains all the information regarding one migration task, such as the migrating VM/VNF (RAM size, dirty page rate, data compression ratio, remaining dirty pages), source and destination hosts, the scheduling window, assigned routings, the current phase of live migration, etc.
The \textit{MigrationPlanner} takes the current migration tasks in the waiting queue as input and calculates the sequence of multiple migrations and sends the result to the \textit{MigrationScheduler}. If there are additional migration tasks arrive, it will calculate the sequence again based on the on-going and waiting to schedule migration tasks.
\textit{MigrationScheduler} takes charge of starting the migration task based on the output of the \textit{MigrationPlanner}. When a migration complete, the \textit{SDNDataCenter} will send the \textit{event 7} (Table \ref{tb: sim-events}) to trigger the scheduler to start new migrations according to the remaining scheduling planning.
With the events of live migration, the Class \textit{SDNDataCenter} emulate the live migration in every phase as shown in Fig. \ref{fig: migration}: (1) checking the availability of network and computing resources; (2) sending the memory and dirty pages to the destination hosts iteratively; (3) checking the current downtime and iterative rounds with the thresholds; (4) pausing the workload processing and refusing the new packets arrive at the instance; (5) resuming the workload processing and rerouting the network packets to the new location. (6) noticing the on-line scheduler about the completion; (7) if selected, storing the statistic for every migration step. The \textit{NetworkOperatingSystem} calculates the routings and allocated bandwidth to the migration flows based on the selected network routing policy and bandwidth sharing scheme. It simulates the network packet transmission based on the bandwidth and delay along the path, packs and unpacks the contents from and to the compute nodes.


\subsection{Evaluation Scenarios and Configurations}
In this section, we list the details of various evaluation scenarios and corresponding setups regarding the physical datacenter topologies, virtual topologies (applications), and workloads.

\begin{figure}[htbp]
	\centering
	\includegraphics[width=0.5\linewidth]{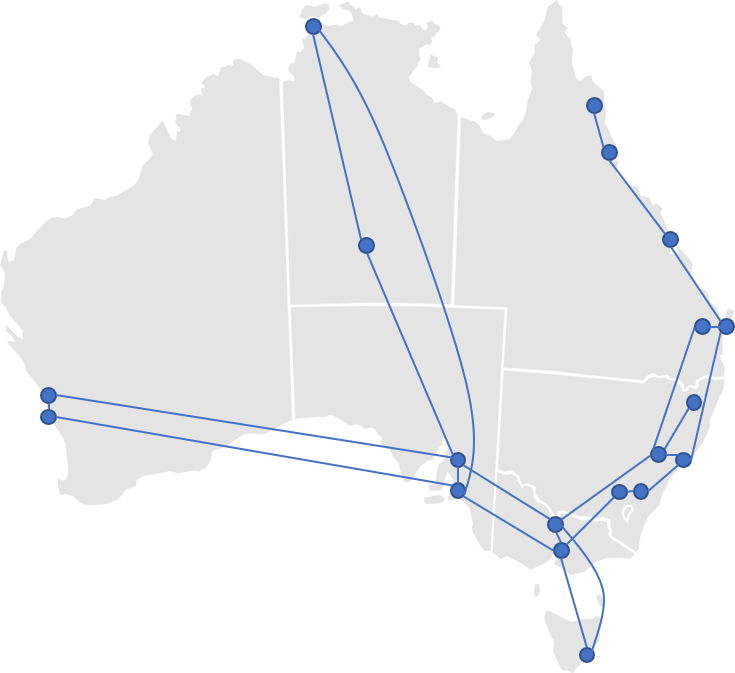}
	\caption{AARNET as the inter-datacenter WAN \cite{knight2011internet}}
	\label{fig: aarnet-map}
\end{figure}

For the physical data center topology, we evaluated the performance of multiple migrations planning algorithms in both (1) WAN environment for Inter-Data Centers Network \cite{knight2011internet} and (2) Intra-Data Center Network (FatTree). 
The three-tier 8-pod FatTree \cite{al2008scalable} intra-data center network consists of 128 physical hosts with the configuration of 24 cores, 10000 MIPS each, 10240 GB RAM, 10PB storage, and 10 Gbps for all physical links. The inter-data center network used in the experiment is shown in Fig. \ref{fig: aarnet-map}. Each link between routers has 10 Gbps bandwidth. Each router as the gateway connects to the local data center cluster through the 40 Gbps link. Each local data center includes 2048 hosts with the same configuration of the one in FatTree which designed to be sufficient for all instances during the experiments. 

\begin{figure}[th]
	\centering
	\begin{subfigure}{0.45\linewidth}
		\includegraphics[width=\linewidth]{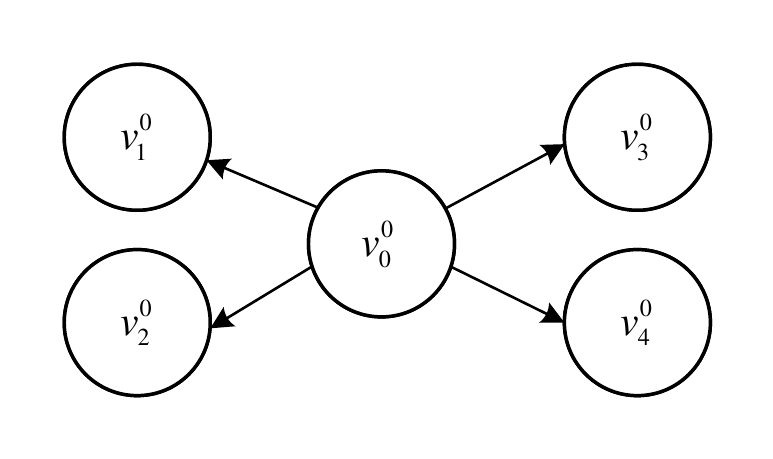}
		\subcaption{\textit{star-to-slave}}
		\label{fig: star-to-slave}
	\end{subfigure}\hfil
	\begin{subfigure}{0.45\linewidth}
		\includegraphics[width=\linewidth]{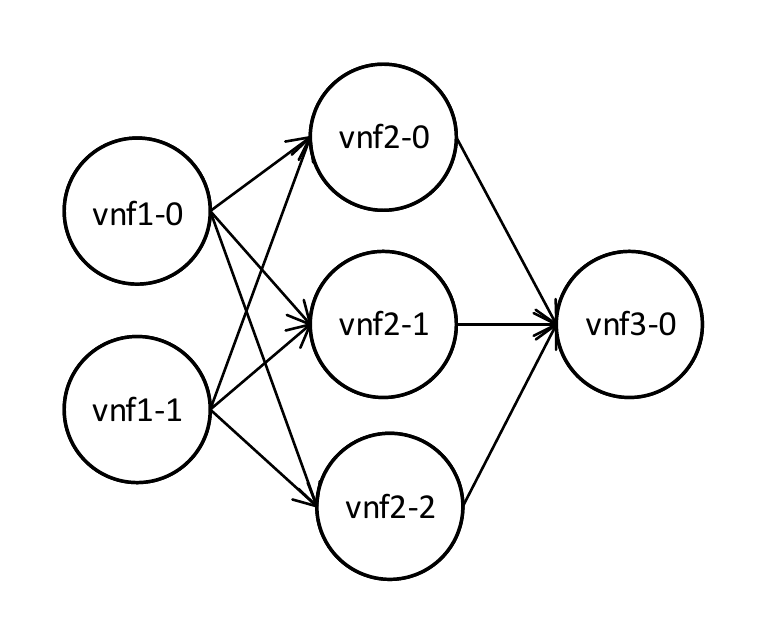}
		\subcaption{\textit{sfc}}
		\label{fig: sfc}
	\end{subfigure}\hfil \\
	\begin{subfigure}{\linewidth}
		\includegraphics[width=\linewidth]{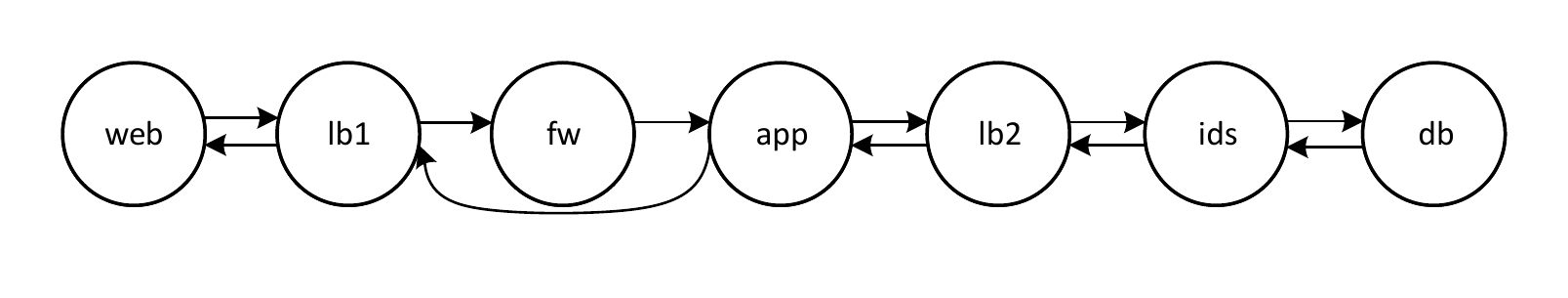}
		\subcaption{\textit{wiki} for 3-tier web application with SFCs}
		\label{fig: web-topo}
	\end{subfigure}\hfil
	\caption{Virtual Topologies used in the experiments}
	\label{fig: virtual-topo}
\end{figure}

Regarding the types of virtual topology (application), we selected them by different flavors and connectivity. Table \ref{tb: flavor} illustrates the flavors we used for different applications, such as multi-tier web applications and SFCs. In general, we generated 4 different types of virtual topologies: (1) \textit{single}; (2) \textit{star-to-slave}; (3) \textit{sfc}; and (4) \textit{wiki} (multi-tier web application with SFCs).

%

There is no connection or network communication between VMs in the \textit{single} topology. 
For every group of \textit{star-to-slave}, there is one master instance that connects to other slave instances in a star fashion. The network requests and workloads are only sent from the master to the slave instance. Figure \ref{fig: star-to-slave} indicates a \textit{star-to-slave} virtual topology where $v_0^0$ is the master instance and $v_1^0$ to $v_4^0$ are the slave instances.
The \textit{sfc} consists of VNFs chained together where each tier can have multiple identical VNFs with the same function. The workloads are sent evenly to the VNFs with same function as shown in Fig. \ref{fig: sfc}.
Each request generated between two VMs/VNFs in \textit{star-to-slave} and \textit{sfc} experiments consists of three parts: computing workload in the sender VM (instruction numbers), data transmission workload (bits), and computing workload in the recipient VM. The request is first processed in the sender VM. Then, network data is generated and sent to the recipient VM. Finally, the recipient VM processes the request.
The service request arrival time of \textit{star-to-slave} and \textit{sfc} experiments are generated in a finite time interval based on the Poisson distribution with a mean as 20 and 200 per second, respectively. 
Each packet size ($pSize$) is generated in the normal distribution with $pSize$ as the mean value and $0.1pSize$ as the variance.  
The CPU processing workloads in the sender and recipient are generated based on the given workload size ($loadsize$) of request sender and recipient in the normal distribution with $loadsize$ as the mean value and $0.2loadsize$ as the variance. 
The $pSize$ of each packet is 5 Mbits. The $loadsize$ for request sender and recipient is 100 and 50, respectively.

In the scenarios of \textit{wiki}, we simulate the three-tier web applications consisting of web (web), application (app), and database (db) servers. 
We generate synthetic workloads based on Wikipedia trace \cite{van2009wikibench} following the three-tier application model \cite{ersoz2007characterizing}.
Network traffics between servers is forwarded to different types of VNFs: Load Balancer (lb), Firewall (fw), and Intrusion Detection System (ids). The configuration of different types of servers and VNFs are shown in Table \ref{tb: flavor} and \ref{tb: setup}. As shown in Figure \ref{fig: web-topo}, flows from the web servers are forwarded to VNF lb1 then fw before reach to the app servers. Meanwhile, flows from the app servers are forwarded to VNF lb2 and ids before reach to the db servers. For those flows coming back to the web servers from db servers, they need through VNFs ids and lb2 then app servers and the VNF lb1. 
In addition to those general VM specifications, VNFs have a specific field named MIPO (million
instructions per operation)\cite{cloudsimsdn-nfv}, which models the throughput of the VNF. The MIPO specifies the CPU workload length for a single network operation provided by a VNF. Thus, it can provide the throughput of the VNF along with the MIPS. For example, a VNF with 10000MIPS and 10MIPO can handle 100 operations (request) per second. We assign MIPO to Load Balancer, IDS, Firewall as 20, 200, and 800, respectively.

\subsection{Results and Analysis}
In this section, we evaluate the performance of our proposed algorithms SLAMIG through several experiments, including \textbf{migration performance}, \textbf{QoS awareness}, \textbf{deadline awareness}, and \textbf{energy consumption}. 
In order to compare with other multiple migration scheduling algorithms \cite{bari2014cqncr, wang2017virtual}, we use the similar simulation settings in terms of initial placement and dynamic resource management policy. Using the settings, we highlight the benefits of our multiple migration planning and scheduling algorithm compared to other algorithms. Note that, given the multiple migration requests provided by the dynamic resource management policies, multiple migration planning and scheduling algorithms are not confined to any specific resource reallocation scenario.
The initial placement of all instances are generated in the way that connected VMs and VNFs are distributed among hosts in Least Full First (LLF) manner. The dynamic resource algorithm generates migration requests to consolidate all connected VMs and VNFs into the same host as compactly as possible, and if not, allocate them to the most full hosts. 
The configuration can simulate a large amount of resource contention between the multiple migration requests for the dynamic resource management to efficiently utilize the cloud resources.
We compare the performance of SLAMIG with the one-by-one migration policy as the baseline and the other two state-of-art algorithms. One algorithm (CQNCR) \cite{bari2014cqncr} migrates VMs by groups. The other is the approximation algorithm (FPTAS). It optimizes the total migration time by maximizing the total assigned network bandwidth to migrations \cite{wang2017virtual}.

\subsubsection{Migration Performance} \label{sec: mig-performance}

In this section, we evaluated the migration performance in terms of total migration time, total downtime, average execution time, total transferred data, and processing time. In experiment \textit{single}, we randomly generated a total of 100 to 1000 instances with flavor from micro to large in the inter-data center topology or Wide Area Network (WAN) (Fig. \ref{fig: aarnet-map}). We use the dirty page factor in the simulation experiments, which is the ratio of the dirty memory rate (bits per seconds) to the total memory of the VM (bits) being migrated. For the scenarios of migrating instances with low and high dirty page rate, we randomly generate the dirty page factor from 0.01 to 0.05 and from 0.01 to 0.15, perspectively. The dirty page rate (Gbps) is the product of the total memory size and the dirty page factor. 

\begin{table}[htbp]
	\centering
	\caption{Experiment scenarios profile of \textit{wiki}} \label{tb: setup}
	\resizebox{\linewidth}{!}{
		\begin{tabular}{|l|l|l|l|l|l|l|l|l||l|l|}
			\hline
			Scen-	 	&	\multicolumn{4}{c|}{VNFs $\#$}	&	\multicolumn{3}{c|}{VMs $\#$}	& Reser. bw & Target Rate	&	 Mig.  \\
			\cline{2-9}
			arios						&	lb1 & fw & lb2 & ids		&	web	&	app	& 	db	&	(Mbps)	&	(Request/s)	&	$\#$		\\
			\hline
			wiki-s1		&	1 & 3 & 1 & 3		&	8	&	24	& 	2	&	2	& 7.8402 & 34	\\
			wiki-s2		&	2 & 6 & 2 & 6		&	32	&	96	& 	8	&	2	& 1.9601 & 118	\\
			wiki-s3		&	2 & 6 & 2 & 6		&	80	&	240	& 	20	&	2	& 1.5680 & 180	\\
			\hline
			
		\end{tabular}
	}
	\label{tb: wiki-exp}
\end{table}

Furthermore, we evaluate the migration performance of \textit{wiki} scenarios in FatTree. Table \ref{tb: wiki-exp} illustrates the details of three scenarios in the \textit{wiki} experiment, including the virtual topologies of SFCs and multi-tier web applications, reserved virtual bandwidth, the request arrival rate, and the number of migration tasks. The dirty page factor is set as 0.001 for all instances.

\paragraph{Single VM Topology in Inter-Data Centers}
\begin{figure}[th]
	\centering
	\begin{subfigure}{0.33\linewidth}
		\includegraphics[width=\linewidth]{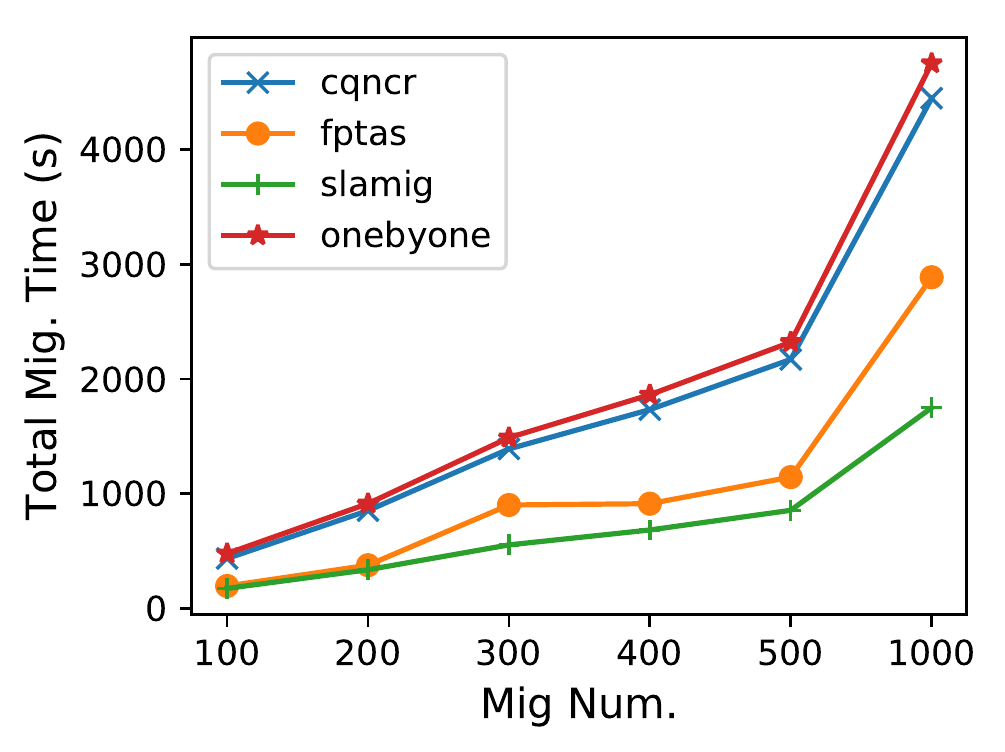}
		\subcaption{total migration time with low dirty rate}
		\label{fig: single1}
	\end{subfigure}\hfil
	\begin{subfigure}{0.33\linewidth}
		\includegraphics[width=\linewidth]{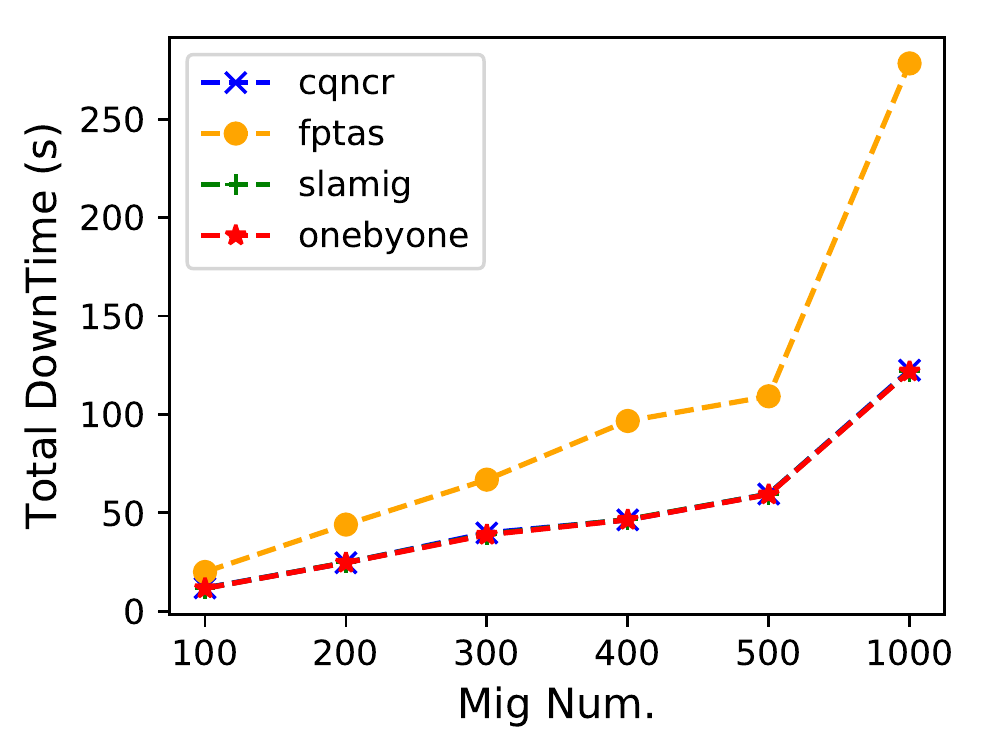}
		\subcaption{total downtime with low dirty rate}
		\label{fig: single2}
	\end{subfigure}\hfil \\
	\begin{subfigure}{0.33\linewidth}
		\includegraphics[width=\linewidth]{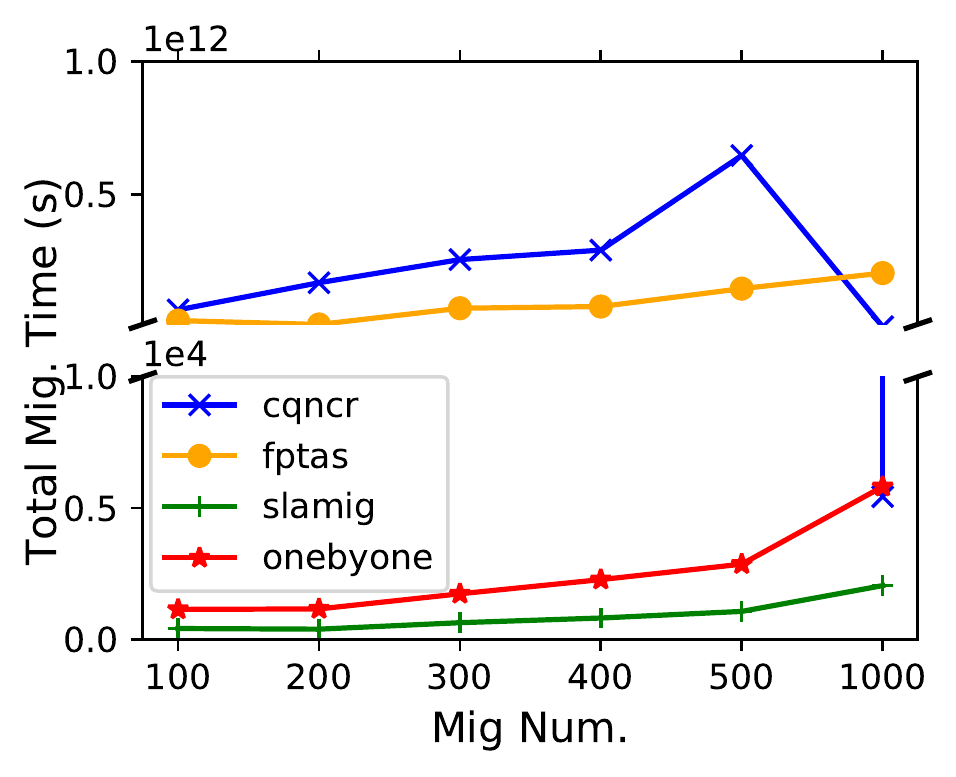}
		\subcaption{total migraion time with high dirty rate}
		\label{fig: single3}
	\end{subfigure}\hfil
	\begin{subfigure}{0.33\linewidth}
		\includegraphics[width=\linewidth]{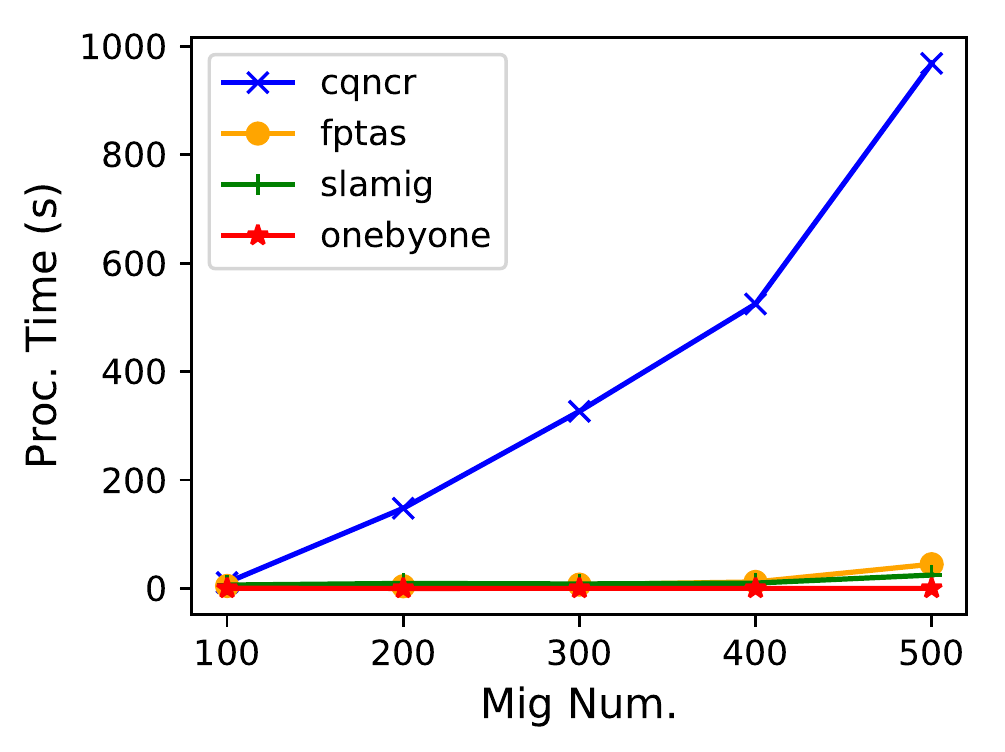}
		\subcaption{processing time with low dirty rate}
		\label{fig: single4}
	\end{subfigure}\hfil
	\caption{Live migration of non-connected VM (\textit{single}) in AARNET}
	\label{fig: single}
\end{figure}

First, we evaluated the migration performance in a large scale manner from 100 to 1000 total migration tasks (Fig. \ref{fig: single}). The results indicate that in SLAMIG can achieve the best total migration time without scarifying the downtime performance in both high dirty page rate and low dirty page rate cases. Regarding the processing time of multiple migration planning, our algorithm is less time consuming than the approximated algorithm (FPTAS) and iterative heuristic grouping (CQNCR). In the scenarios of the low-dirty-page-rate single experiment, the total migration time of SLAMIG is $62.85\%$ to $63.69\%$ less than the baseline and $10.50\%$ to $39.41\%$ less than the FPTAS. By starting the migrations group by group at the same time, the total migration time of CQNCR is only marginally smaller than the result of the baseline (maximum $9.06\%$). Meanwhile, as shown in Fig \ref{fig: single2}, the total downtime of SLAMIG is at least $40.27\%$ and at most $55.87\%$ less than the FPTAS. 

For the result of algorithm running time, we observe that the SLAMIG algorithm can significantly reduce the computation time compared to solving the approximate MIP problem in FPTAS. In addition to Fig. \ref{fig: single4}, when there are 1000 migration tasks, the processing times of CQNCR, FPTAS, SLAMIG are 15471.29, 89.94, and 30.23 seconds respectively. When performing 500 migration tasks, the processing time of SLAMIG (24.64s) is $44.70\%$ less than that of FPTAS. The runtime of sequential scheduling is less than 1 second, because in our experiments, the available sequence only needs to be calculated once as all sequential combinations are schedulable. For the algorithm CQNCR, after updating the network bandwidth and computing resources in each round, it iteratively groups the migrations in a greedy manner and selects the migration group with the most positive impact. Thus, when the number of migration tasks increases, the processing time will increase dramatically (Fig. \ref{fig: single4}). Compared with CQNCR, our proposed algorithm can calculate all concurrent migration groups in one round. Since each migration task has been given a weight in the dependency graph, we also generate the largest possible migration group with minimal weight. Therefore, it can achieve better performance in total migration time. Note that in our algorithm, generating a migration dependency graph takes up most of the processing time in multiple routing environments. For the single routing environment such as FatTree, we only need to check the source and destination hosts, which will further reduce the processing time.

Fig. \ref{fig: single3} shows the details of the experiment of single instances with a high dirty rate. Compared with the other two algorithms, SLAMIG can maintain the performance of the total migration time. By allowing other migration tasks to be initiated when there is a small amount of bandwidth to maximize the overall network transmission rate, FPTAS may cause significant performance degradation in both total migration time and downtime. In the worst case, the total migration time shown is even greater ($10^6$ times) than the result of one-by-one scheduling. Moreover, all migration start times are based on the prediction model in CQNCR. Inevitably, in the worst case, several migrations will start when their resource-dependent migration tasks haven not been completed, which will cause the allocated bandwidth to be less than the dirty page rate. In other words, the allocated bandwidth is insufficient to converge the migration in the worst case.

\paragraph{Web Application Topology in FatTree}
\begin{figure}[th]
	\centering
	\begin{subfigure}{0.33\linewidth}
		\includegraphics[width=\linewidth]{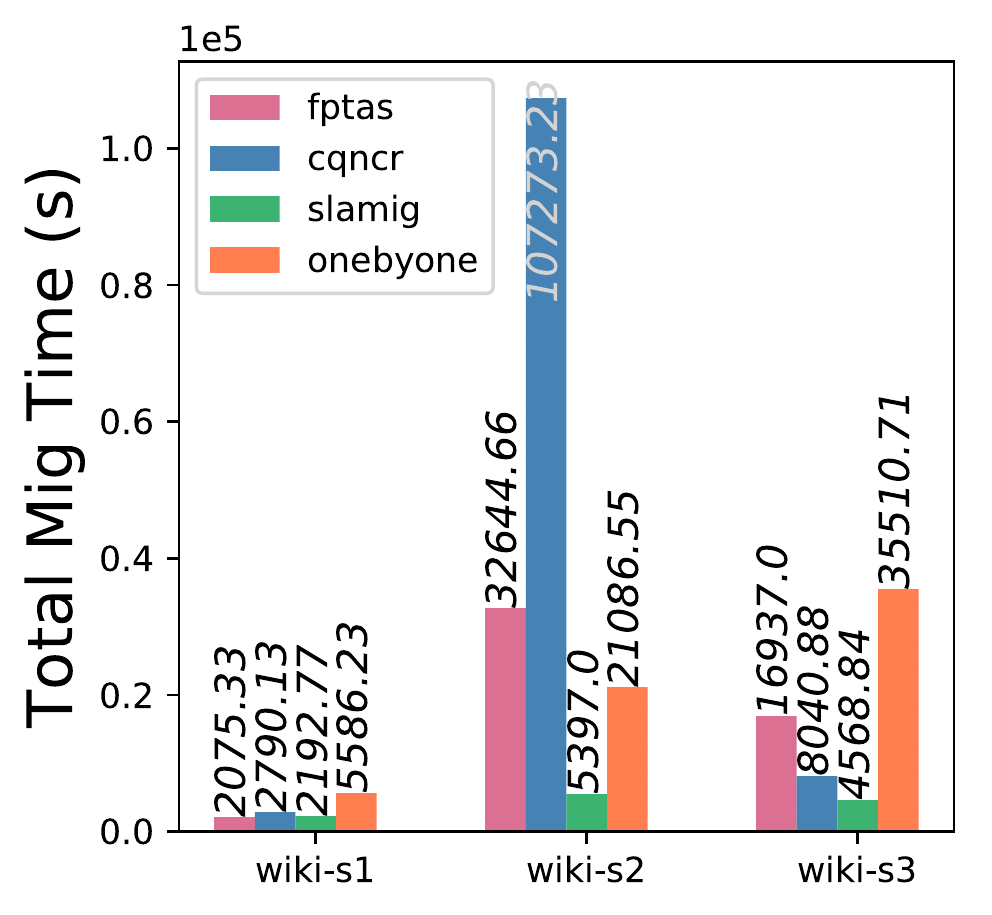}
		\subcaption{total migration time}
		\label{fig: wiki1}
	\end{subfigure}\hfil
	\begin{subfigure}{0.33\linewidth}
		\includegraphics[width=\linewidth]{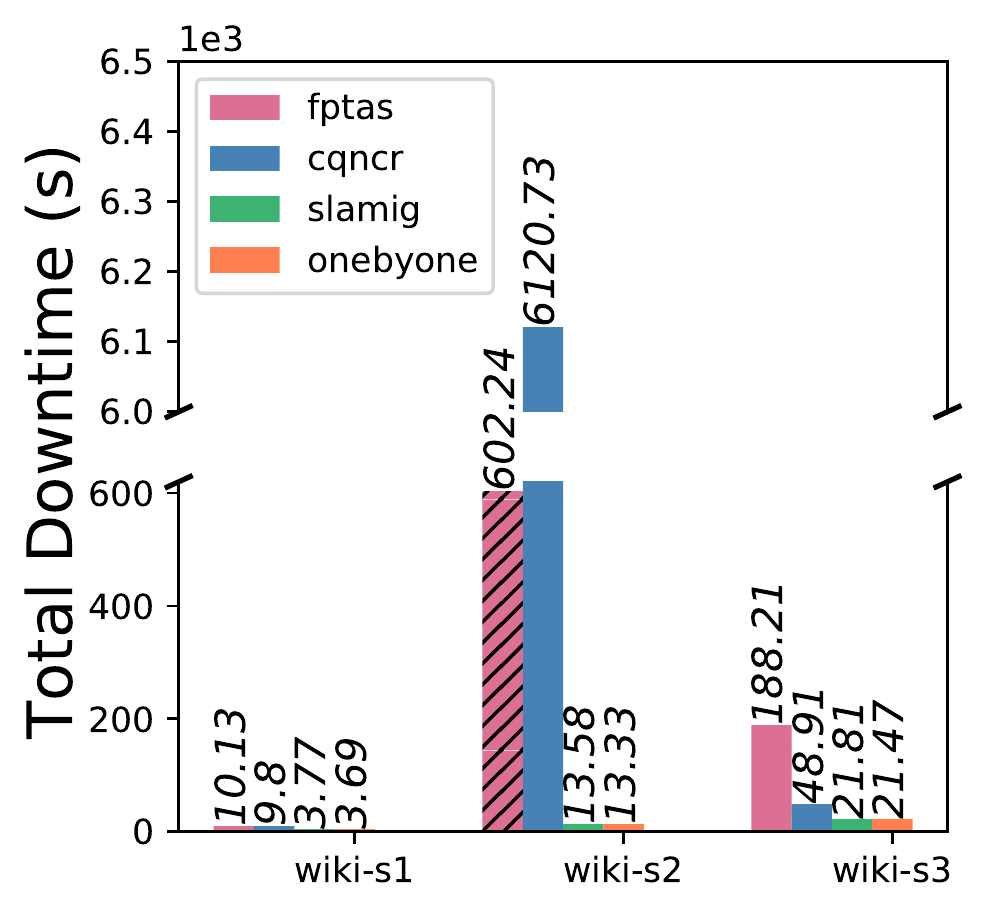}
		\subcaption{total downtime}
		\label{fig: wiki2}
	\end{subfigure}\hfil \\
	\begin{subfigure}{0.33\linewidth}
		\includegraphics[width=\linewidth]{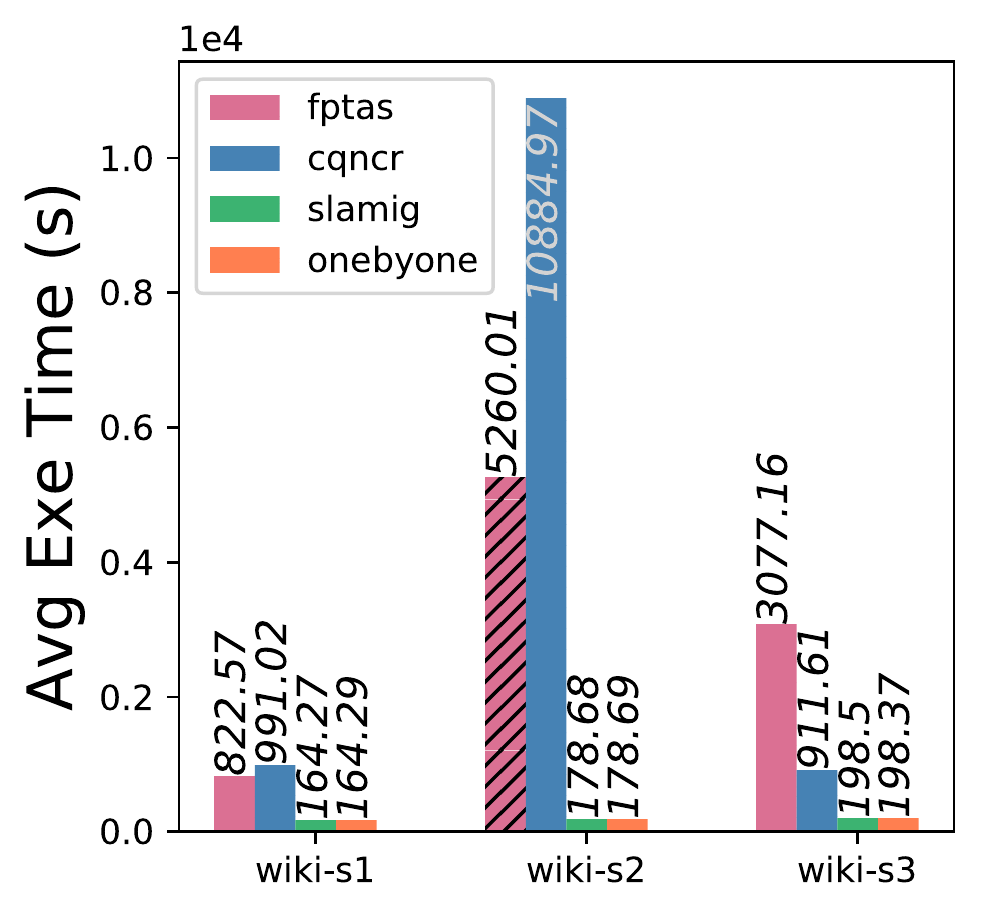}
		\subcaption{average execution time}
		\label{fig: wiki3}
	\end{subfigure}\hfil
	\begin{subfigure}{0.33\linewidth}
		\includegraphics[width=\linewidth]{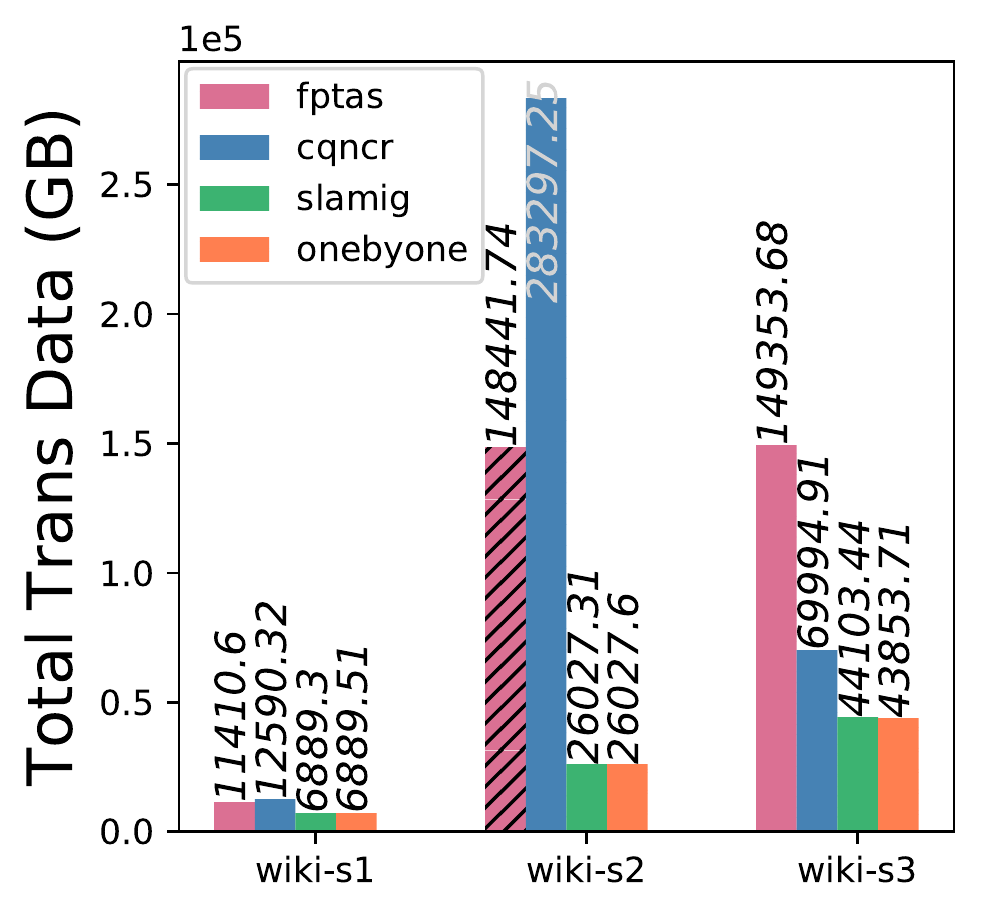}
		\subcaption{total transferred data}
		\label{fig: wiki4}
	\end{subfigure}\hfil
	\caption{Live migration of multi-tier applications with SFCs (\textit{wiki}) in FatTree}
	\label{fig: wiki}
\end{figure}

In the experiment of \textit{wiki}, we evaluated the algorithm performance regarding the total migration time, total downtime, average execution time, and total transferred data during the live migrations (Fig. \ref{fig: wiki}). In all three scenarios, the SLAMIG achieves the optimal total migration time while maintaining other migration performance criteria at the level of sequential scheduling. Compared with the baseline, the SLAMIG reduce the total migration time by $60.74\%$, $74.41\%$, and $87.13\%$. The results are $-5.66\%$, $83.47\%$, and $73.02\%$ less than FPTAS and $21.41\%$, $94.96\%$, and $43.17\%$ less than CQNCR.

In some cases, such as wiki-s1 in Fig. \ref{fig: wiki1}, we noticed that the total migration time of FPTAS may be slightly better than our algorithm. It is because several migration tasks can be scheduled in the same paths when a small amount of bandwidth is available to maximize the overall network transmission rate. One migration can be started even the allocated bandwidth is smaller than the dirty page rate. Although the sum of migration execution time is larger, the total migration time may be smaller due to the early start time. As mentioned in Section \ref{sec: impact-bw-dirty-rate}, we argue that it will increase the average execution time of each migration task (Fig. \ref{fig: wiki3}), resulting in larger downtime (Fig. \ref{fig: wiki2} and \ref{fig: single2}) and total transferred data (Fig. \ref{fig: single4}).

Considering total downtime, average execution time, and total transferred data, we should concurrently schedule the resource-dependent migration tasks to alleviate the impact of multiple live migrations on the system and guarantee the QoS of the migrating instances. The results indicate that there is no statistical difference between SLAMIG and the sequential scheduling in these parameters. However, the FPTAS and CQNCR drastically increase the total downtime by 1.75/1.66, 44.19/458.23, 7.77/1.28 times, the average execution time by 4.006/5.03 times, 28.44/59.92, 14.51/3.60, and the total transferred date by 0.66/0.83, 4.70/9.88, 2.41/0.60 times, respectively. Although the FPTAS and CQNCR can achieve a better performance of total migration time compared to the baseline in other scenarios, bandwidth sharing among resource-dependent instances with large memory and high dirty page rate will lead to unacceptable results (wiki-s2).

\paragraph{Summary}
(1) SLAMIG achieves the optimal migration performance in terms of the total migration time, downtime, average execution time, and total transferred data, while the processing time is less than the CQNCR iterative grouping and FPTAS approximation algorithm.
(2) The prediction model of migration is used to estimate the execution time of one migration and the total migration time of a concurrent migration group. However, it is not efficient to assign a fixed start time for a live migration only based on the prediction model. In an independent migration group, the execution time varies, which leads to multiple time gaps between the completion time and the estimated start time of the next group. Moreover, in the real environment, the real-time dirty page rate may be different from the historical statistics and the current monitoring value. In a dynamic network environment, the available network bandwidth used in the prediction model may also change over time. In short, the prediction execution time is not necessarily identical to the actual time during the scheduling, which will cause two resource-dependent migrations to run concurrently. Therefore, it is essential for the on-line scheduler to dynamically schedule migration tasks according to the plan. 
(3) By maximizing the total network transmission rate, the total migration time can be reduced to a certain extent, but the optimal migration performance cannot be achieved. If one migration starts with the allocated bandwidth below its dirty page rate, it will extremely enlarge the execution time. For migrations with large dirty page rates, allocating bandwidth that is just slightly larger than the dirty page rate will still result in an unacceptable migration performance with a large downtime and memory-copy iteration round. 
Therefore, we should not neglect the concurrency or resource sharing dependencies between different migration tasks.
(4) Regarding the performance and impact of multiple migration scheduling, total migration time is not the only parameter that needs to be optimized.
The average bandwidth for each migration can also reflect the efficiency of multiple migration scheduling.
A larger allocated bandwidth means smaller single migration execution and the downtime. As shown in the Equation \ref{eq:round-number}, it will also result in fewer iteration rounds for dirty page copying. Thus, it we should also achieve better performance in terms of average bandwidth of each migration resulting in better average execution time, total/average migration downtime, and total/average transferred data.

\subsubsection{QoS-Aware}
\begin{figure}[th]
	\centering
	\begin{subfigure}{0.33\linewidth}
		\includegraphics[width=\linewidth]{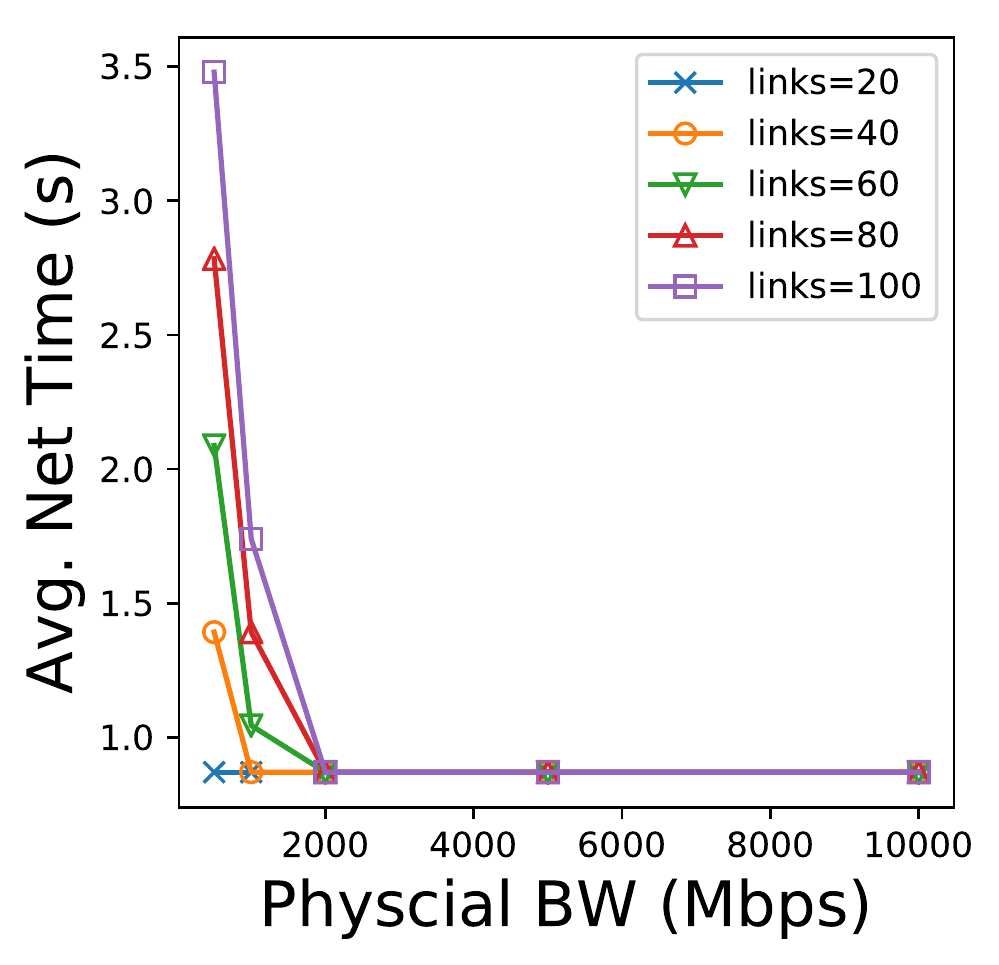}
		\subcaption{20Mbps reserved}
		\label{fig: factor1}
	\end{subfigure}\hfil
	\begin{subfigure}{0.33\linewidth}
		\includegraphics[width=\linewidth]{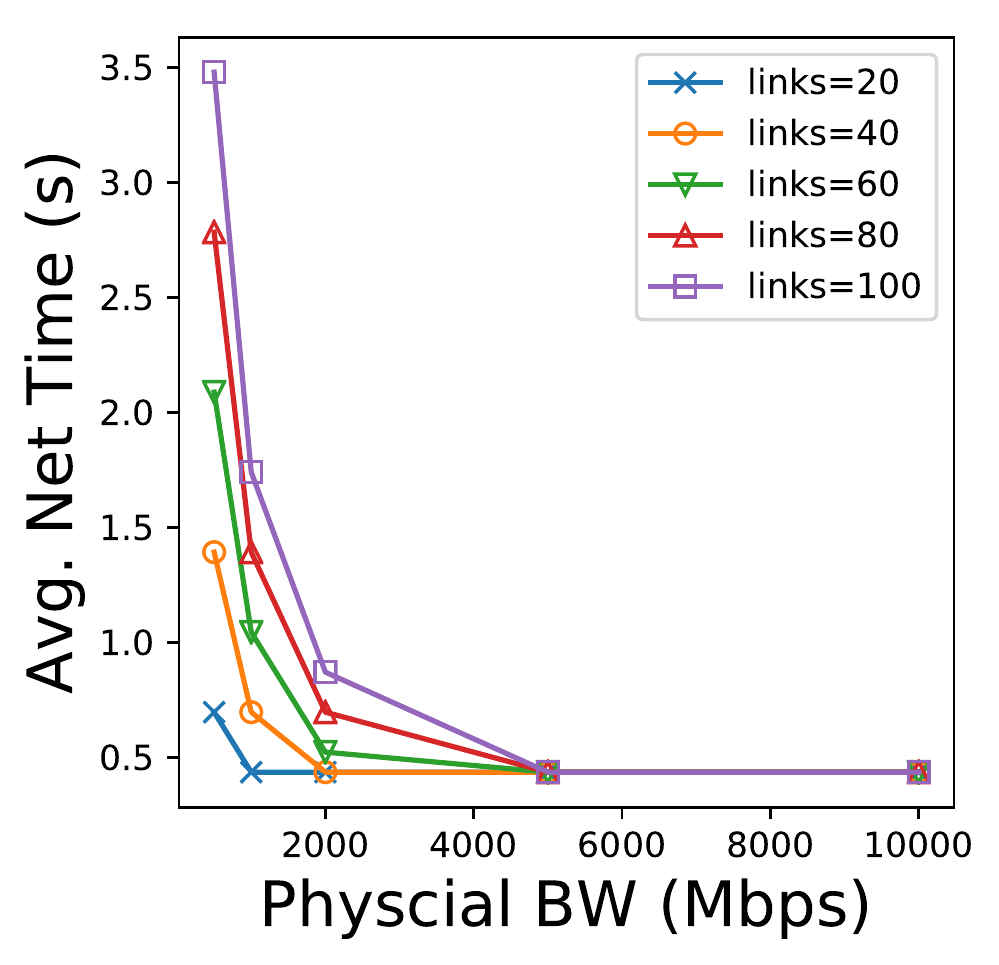}
		\subcaption{40Mbps reserved}
		\label{fig: factor2}
	\end{subfigure}\hfil
	\begin{subfigure}{0.33\linewidth}
		\includegraphics[width=\linewidth]{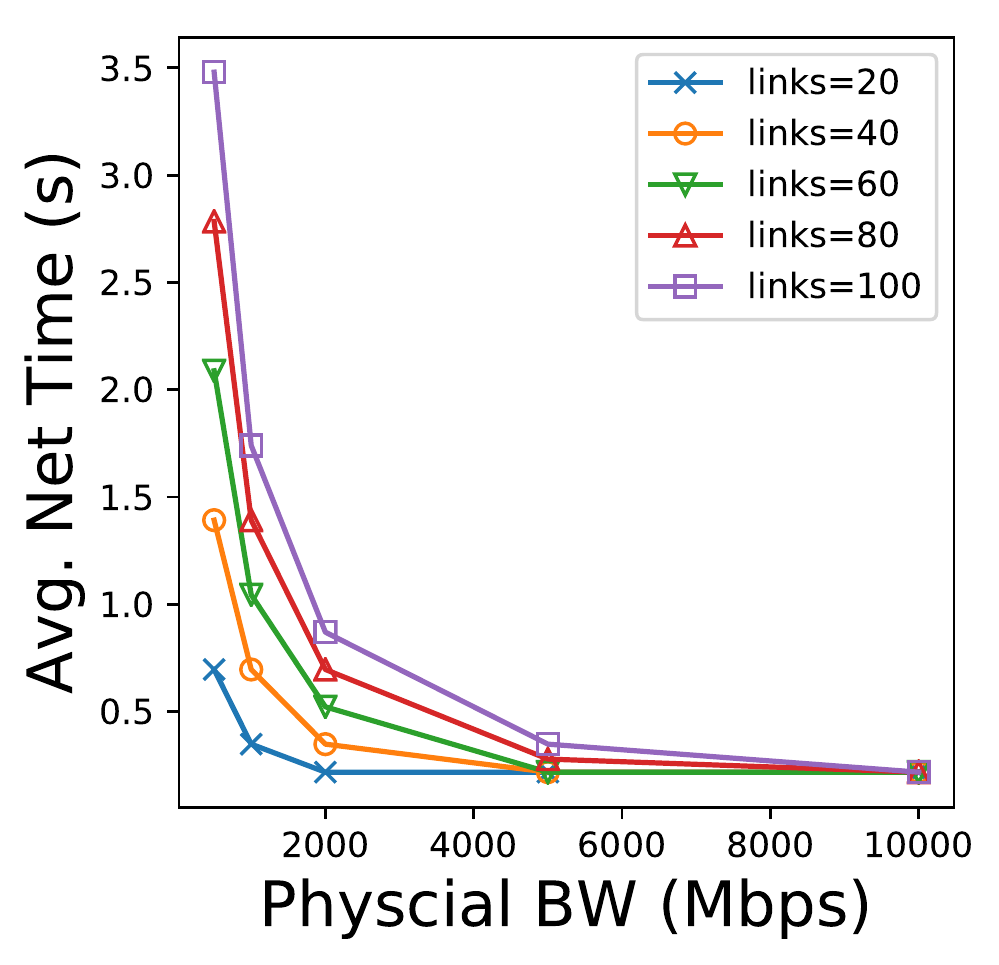}
		\subcaption{80Mbps reserved}
		\label{fig: factor4}
	\end{subfigure}\hfil
	\caption{Average network transmission time without live migration with different number of links and reserved virtual link bandwidth under \textit{ratio} bandwidth sharing policy}
	\label{fig: links}
\end{figure}

\begin{figure}[th]
	\centering
	\begin{subfigure}{0.33\linewidth}
		\includegraphics[width=\linewidth]{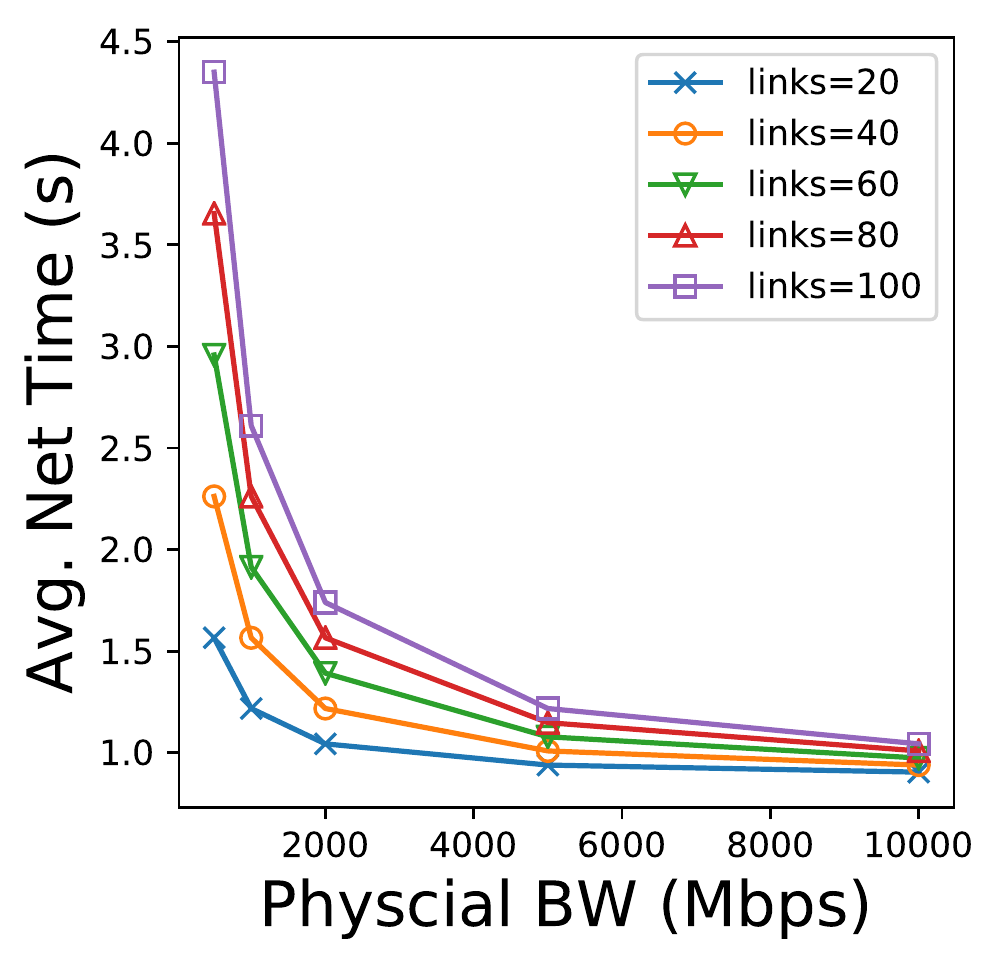}
		\subcaption{20Mbps reserved}
		\label{fig: sharing1}
	\end{subfigure}\hfil
	\begin{subfigure}{0.33\linewidth}
		\includegraphics[width=\linewidth]{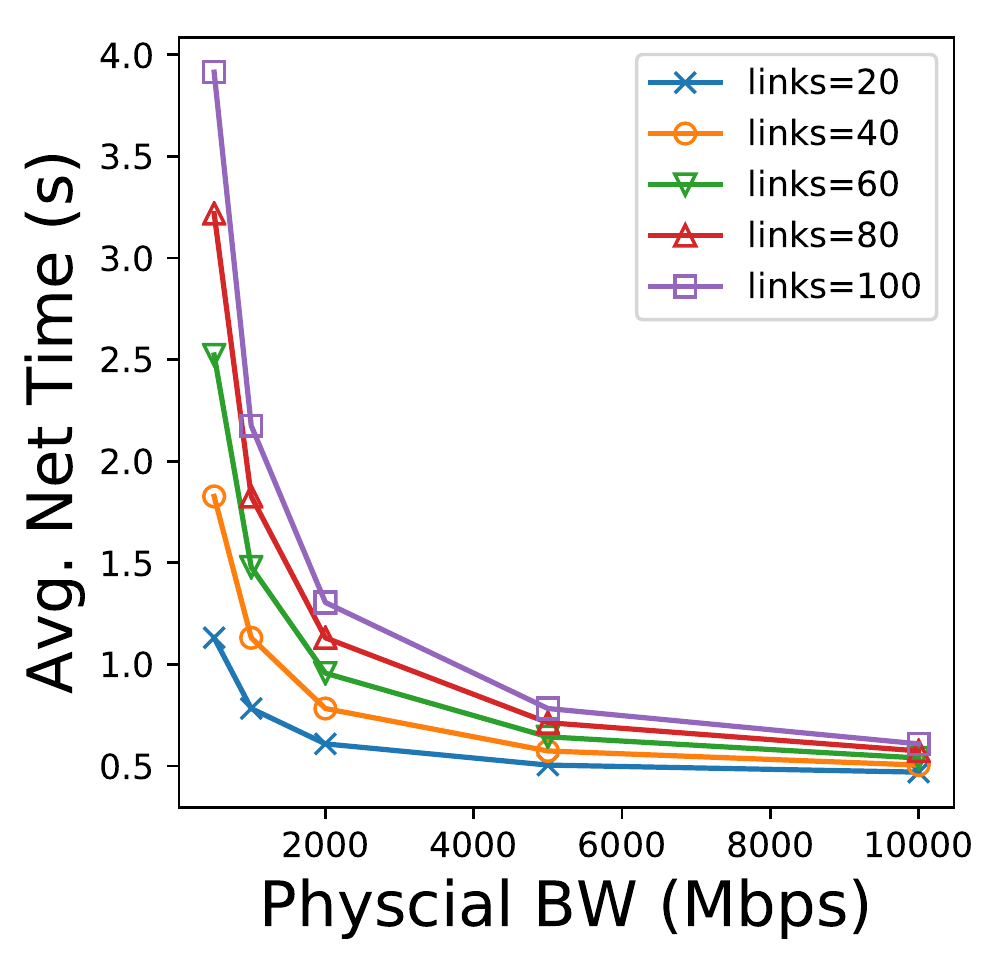}
		\subcaption{40Mbps reserved}
		\label{fig: sharing2}
	\end{subfigure}\hfil 
	\begin{subfigure}{0.33\linewidth}
		\includegraphics[width=\linewidth]{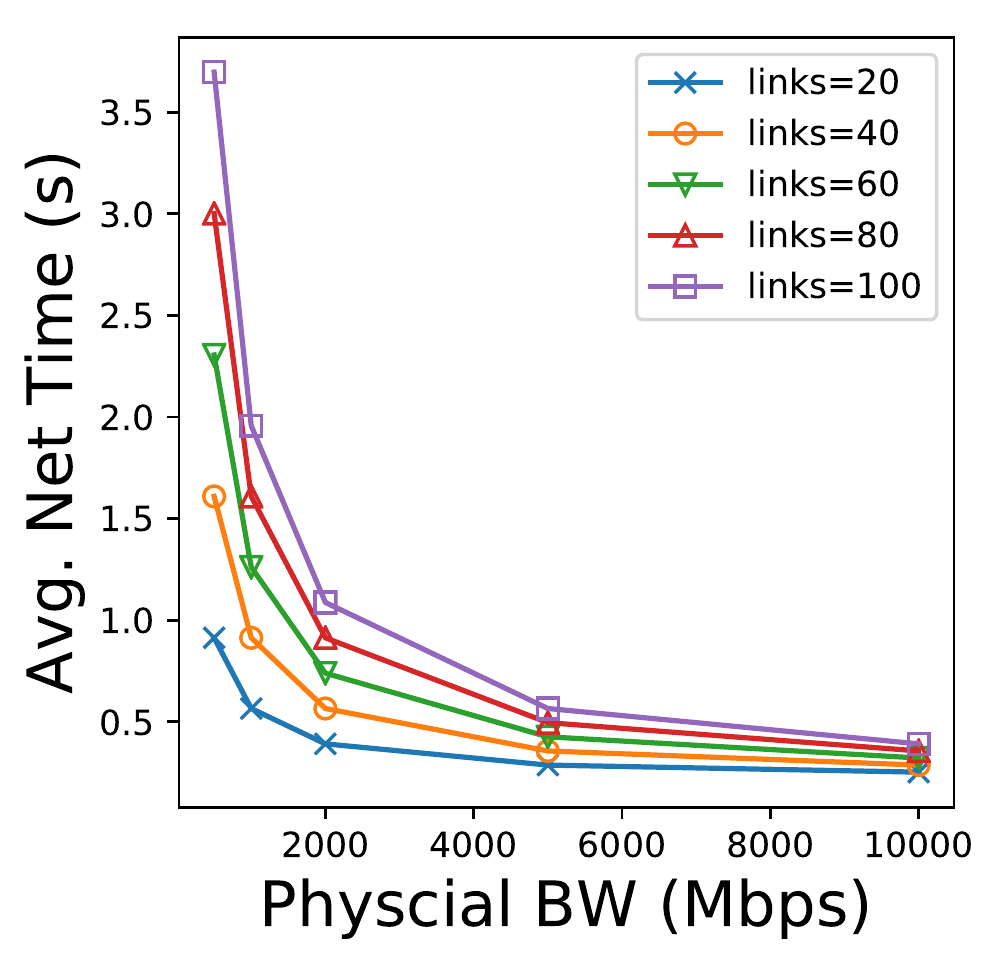}
		\subcaption{80Mbps reserved}
		\label{fig: sharing4}
	\end{subfigure}\hfil
	\caption{Average network transmission time during live migration with different number of links and reserved virtual link bandwidth under \textit{ratio} bandwidth sharing policy}
	\label{fig: sharing}
\end{figure}

In this experiment, we evaluated the impact of multiple migration planning on QoS in terms of the network transmission time of application requests.

There are three network bandwidth sharing policies to manage the migration flow in Section \ref{sec: mig-performance}: (1) \textit{free} used by FPTAS; (2) \textit{reserved} used by CQNCR; and (3) \textit{ratio} used by SLAMIG and OneByOne. 	
The bandwidth sharing solutions proposed by FPTAS and CQNCR which do not consider the bandwidth competition can only be adopted in an ideal scenario where the remaining bandwidth for the live migration is sufficient.
For the \textit{free} policy, the live migration can only utilize the available bandwidth along the network paths left by other service traffic. For the \textit{reserved} policy, the live migration only use the remaining unreserved bandwidth left by other virtual links. The available bandwidth reserved by other services can not be allocated to the migration flow. Therefore, under the {free} or {reserved} bandwidth sharing policy, the live migration flow will not affect the network transmission time of other services in terms of network bandwidth competition. 
Note that in the separated control (migration) network \cite{tsakalozos2017live}, the migration flow will not affect the bandwidth allocation of service traffic.
However, we argue that the \textit{free} and \textit{reserved} policies can only be adopted when the remaining bandwidth for the live migration is sufficient to converge the live migration in time. Furthermore, in some worst cases, as shown in Fig. \ref{fig: wiki}, the massive downtime caused by the \textit{free} or \textit{reserved} policy will seriously affect the request response time of the migrating service.

When other service traffic and migration flows compete on the network bandwidth, studies \cite{clark2005live, he2019performance} show the effect of single live migration on the service response time of the migrating VM. Research \cite{he2019performance} also evaluates the impact on the TCP and UDP traffic and \cite{xu2014, deshpande2017} investigate the effect on other service traffic during the migration.
For the \textit{ratio} policy, the actual allocated bandwidth of a network flow is based on the ratio of the reserved bandwidth of the flow to the total bandwidth demand along the network path. It is practical to use \textit{ratio} bandwidth sharing policy when the remaining bandwidth for the migration flow is insufficient to converge the migration or it is urgent to finish the migration to avoid QoS degradation and SLA violations. 

With the \textit{ratio} policy, we first explain the principle of the impact of live migration on the network traffic between VMs. In the experiment, we control the number of virtual links between VMs along the network path of one migration. The network traffic between two VMs is generated based on the \textit{wiki} workload. Figure \ref{fig: links} illustrates the average network transmission time of network traffic between VMs, where reserved bandwidth size for each virtual link, the total number of virtual links in the evaluating network path, and the available bandwidth of the evaluating network path are controlled variables. 
The results indicate that when the total bandwidth of reserved virtual link is lower than the physical bandwidth, the reserved bandwidth of each virtual link can be satisfied. As the number of links increases, the actual bandwidth allocated for each virtual link decreases, which leads to the longer network transmission time.
Figure \ref{fig: sharing} shows the average network transmission time when the service traffic is sharing the bandwidth with one live migration under the \textit{ratio} policy. In our experiments, the reserved bandwidth for live migration is equal to the physical network bandwidth. 
As the physical network bandwidth increases, the impact of live migration on the network transmission time of other service traffic decreases.
Furthermore, it also indicates that when the number of virtual links along the migration path or the reserved bandwidth for each virtual link increases, the live migration has less impact on the network transmission time of service traffic.

\begin{table}[thbp]
	\centering
	\caption{Simulation configurations of \textit{star-to-slave} and \textit{sfc} experiments}
	\resizebox{\linewidth}{!}{
		\begin{tabular}{|l||l|l|l|l||l|l|l|l|}
			\hline
			group & startoslave &   vm	&  link bw (Mbps) & Mig $\#$	& sfc  & vnf   & link bw (Gbps)  & Mig $\#$\\
			\hline
			5	& star-s1		&	25	&	100 		& 19 & 	sfc-s1	 &  21    & 1.0  &  19 \\
			10	& star-s2		&	50	&	100 		& 37 & 	sfc-s2	  &	 43   &	 1.0 &  40 \\
			15	& star-s3		&	75	&	100 		& 55 & 	sfc-s3	  &	  69  &	 1.0 &  65 \\
			\hline
		\end{tabular}
	}
	\label{tb: qos-simulation}
\end{table}

\begin{figure}[th]
	\centering
	\begin{subfigure}{0.33\linewidth}
		\includegraphics[width=\linewidth]{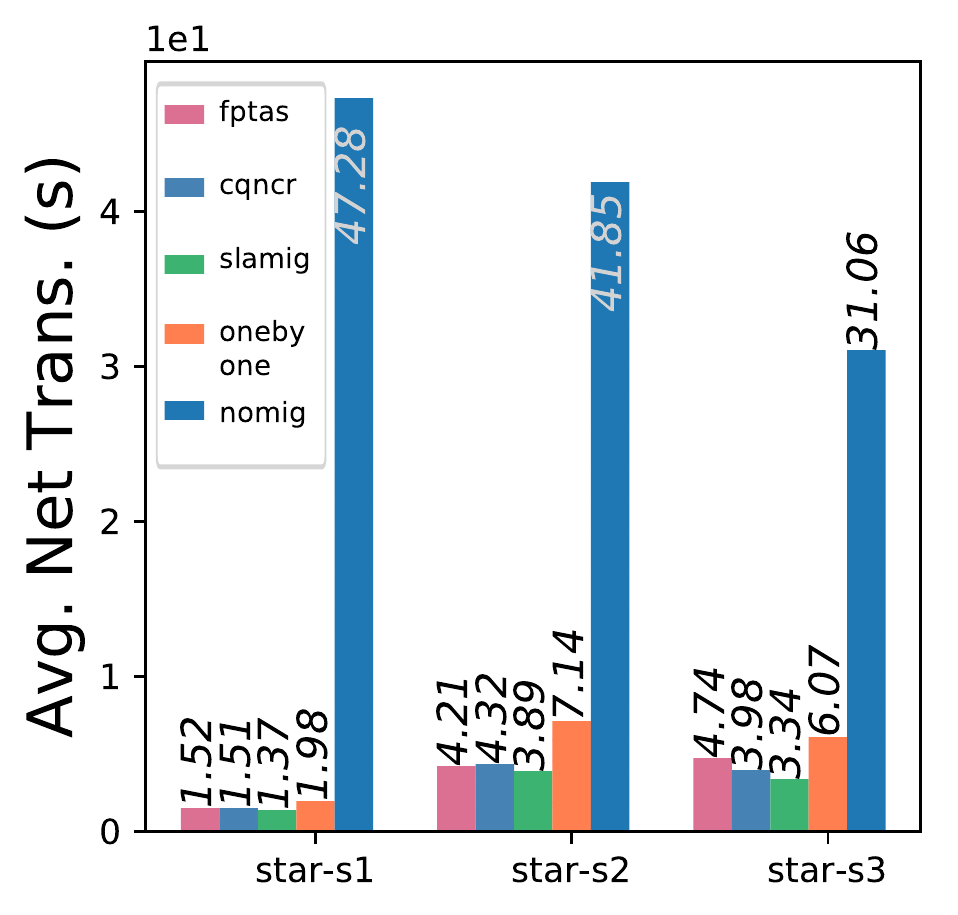}
		\subcaption{\textit{star-to-slave} in WAN}
		\label{fig: qos1}
	\end{subfigure}\hfil
	\begin{subfigure}{0.33\linewidth}
		\includegraphics[width=\linewidth]{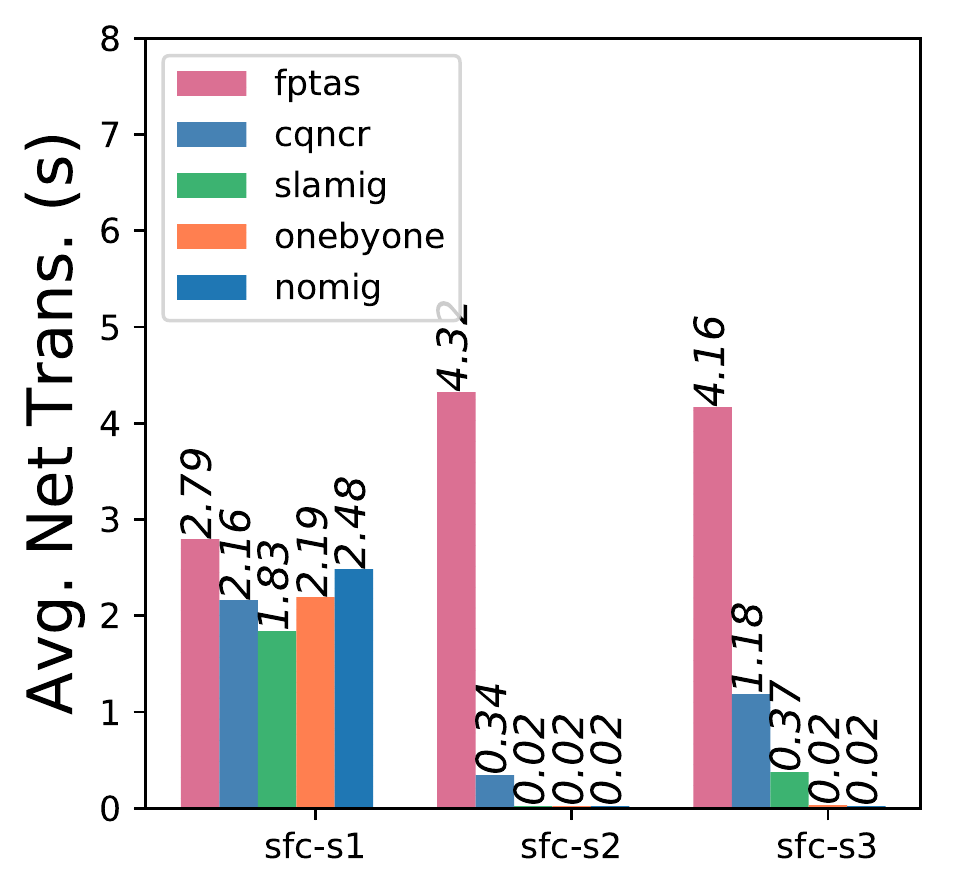}
		\subcaption{\textit{sfc} in WAN}
		\label{fig: qos2}
	\end{subfigure}\hfil
	\begin{subfigure}{0.33\linewidth}
		\includegraphics[width=\linewidth]{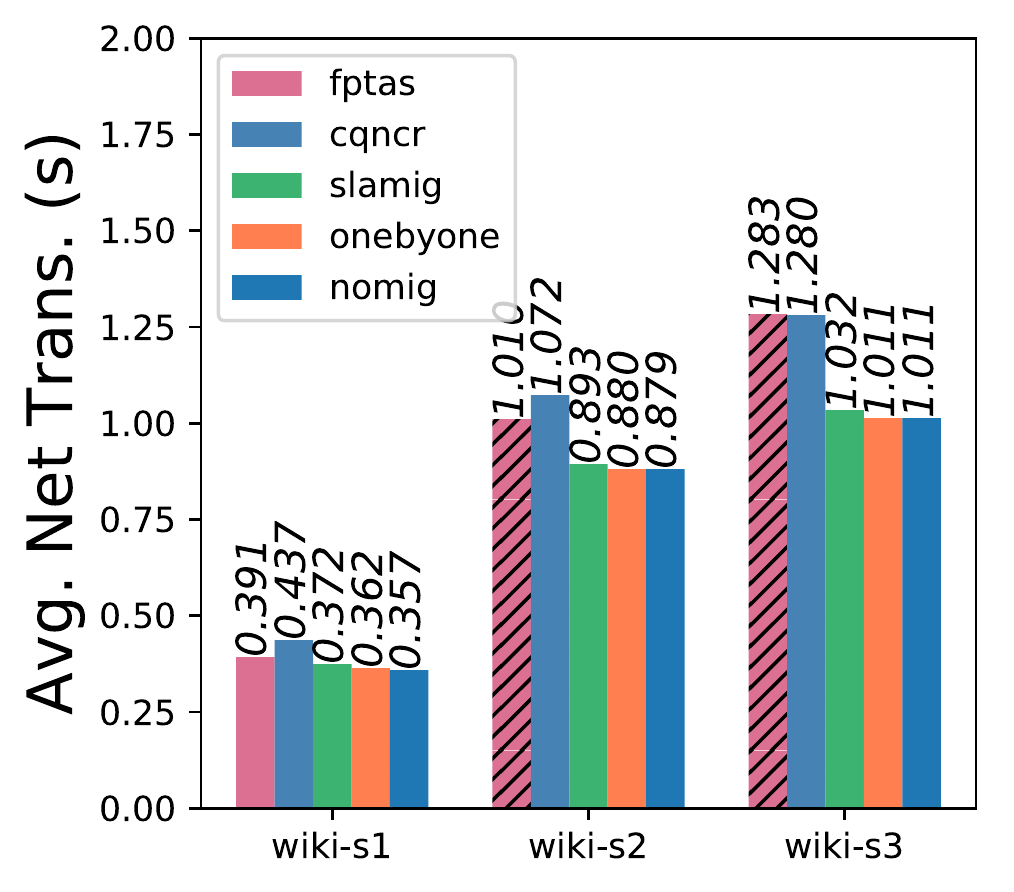}
		\subcaption{\textit{wiki} in FatTree}
		\label{fig: qos3}
	\end{subfigure}\hfil
	\caption{Average network transmission of application requests under \textit{ratio} policy}
	\label{fig: qos}
\end{figure}

To demonstrate the performance of different migration scheduling algorithms with \textit{ratio} bandwidth sharing policy,
in addition to the \textit{wiki} experiment configuration in the FatTree data center network, we also added the experimental results from two types of virtual topologies: (1) \textit{start-to-slave} and (2) \textit{sfc}. The \textit{star-to-slave} and \textit{sfc} experiments are both evaluated in the inter-data center network.
Table \ref{tb: qos-simulation} describes the configuration of the group number, instance number, link reserved bandwidth, and the number of migration tasks in these two experiments. 
We set up the network resources in the way that network traffic within the host can take full advantage of the reserved bandwidth of the virtual link between VMs/VNFs. 
For the master instance with small flavor, the dirty page factor is 0.12, and for the slave instance with tiny flavor and a VNF with large flavor, the dirty page factor is 0.02. 

Fig. \ref{fig: qos1} and \ref{fig: qos2} demonstrate the average network transmission time of applications in the initial placement (nomig): (1) In the \textit{star-to-slave} experiment, applications experience large delay from master to slave instances; (2) In \textit{sfc} experiment, the network transmission time between applications is small in the initial placement. The average network transmission time is 2.48s, 0.02s, and 0.02s, respectively. 

The results of \textit{star-to-slave} indicate that the consolidating migrations can efficiently reduce the delay encountered by the application. The SLAMIG achieves the minimal average network transmission time of application requests in all three scenarios which are 0.14s, 0.32s, and 0.64s less than the second-best results. 
Compared to the non-migration situation, it can also reduce the network transmission time by $95.79\%$, $90.70\%$, and $89.25\%$. 
In the experiment of \textit{sfc}, FPTAS excessively increases the network transmission time of application requests. As the FPTAS algorithm intends to maximize the network transmission rate of all migration tasks, it significantly reduces the transmission bandwidth among the application servers. In scenario sfc-s1, SLAMIG reduces the average network transmission time due to consolidation. Because less total migration time and average execution time will result in a shorter network transmission time during the multiple migrations. For the scenario sfc-s2 and sfc-s3, the initial placement is sufficient to provide enough bandwidth according to the virtual link reservation. SLAMIG does not increase network transmission time in sfc-s2, and only increases 0.35s in sfc-s3, which can guarantee the QoS during the multiple live migrations. For the experiment of \textit{wiki}, SLAMIG can maintain the QoS at the same level of the sequential scheduling with \textit{ratio} bandwidth sharing policy. However, the average transmission time of all service requests increases by $0.04s$,  $0.131s$, and $0.272s$ in FPTAS and  $0.08s$, $0.193s$, and $0.269s$ in CQNCR.

\paragraph{Summary}
Although the migration downtime is an important parameter to evaluate the impact of migration on the migrating instances, the QoS of other services in the data center network is largely ignored \cite{he2019performance}. By utilizing the \textit{free} and {reserved} bandwidth sharing policy, the transmission time of application requests will not be affected. However, in the case where \textit{ratio} bandwidth sharing policy is required to converge the migration, our proposed algorithms can minimize the impact of multiple migrations on the application, thereby ensuring the QoS and mitigating SLA violations.

\subsubsection{Deadline-Aware}
\begin{table}[htbp]
	\centering
	\caption{Evaluation scenarios of deadline-related migrations}
	\begin{tabular}{|l||l|l|l|l|l|}
		\hline
		name		& vm	 & nfv	& D(star) (s)	& D(sfc) (s)  & total mig     \\
		\hline
		star-sfc-5	& 25	 &	23	&	100 		& 	300		  &	46			  \\
		star-sfc-10	& 50	&	40	&	200 		& 	500		  &	87			  \\
		star-sfc-15	& 75	&	75	&	300 		& 	800		  &	138			  \\
		\hline
	\end{tabular}
	\label{tb: deadline-simulation}
\end{table}

\begin{figure}[th]
	\centering
	\begin{subfigure}{0.45\linewidth}
		\includegraphics[width=\linewidth]{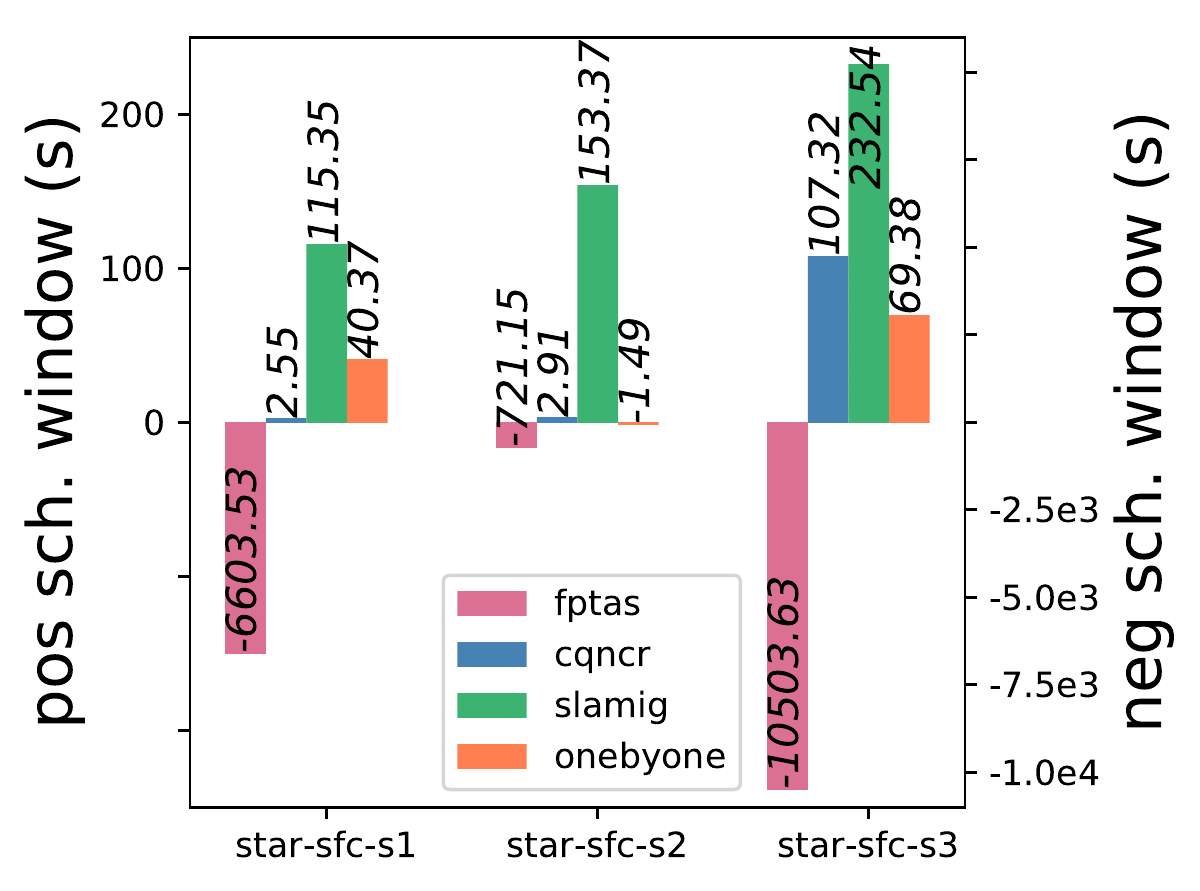}
		\subcaption{avg. remaining scheduling window}
		\label{fig: deadline1}
	\end{subfigure}\hfil
	\begin{subfigure}{0.4\linewidth}
		\includegraphics[width=\linewidth]{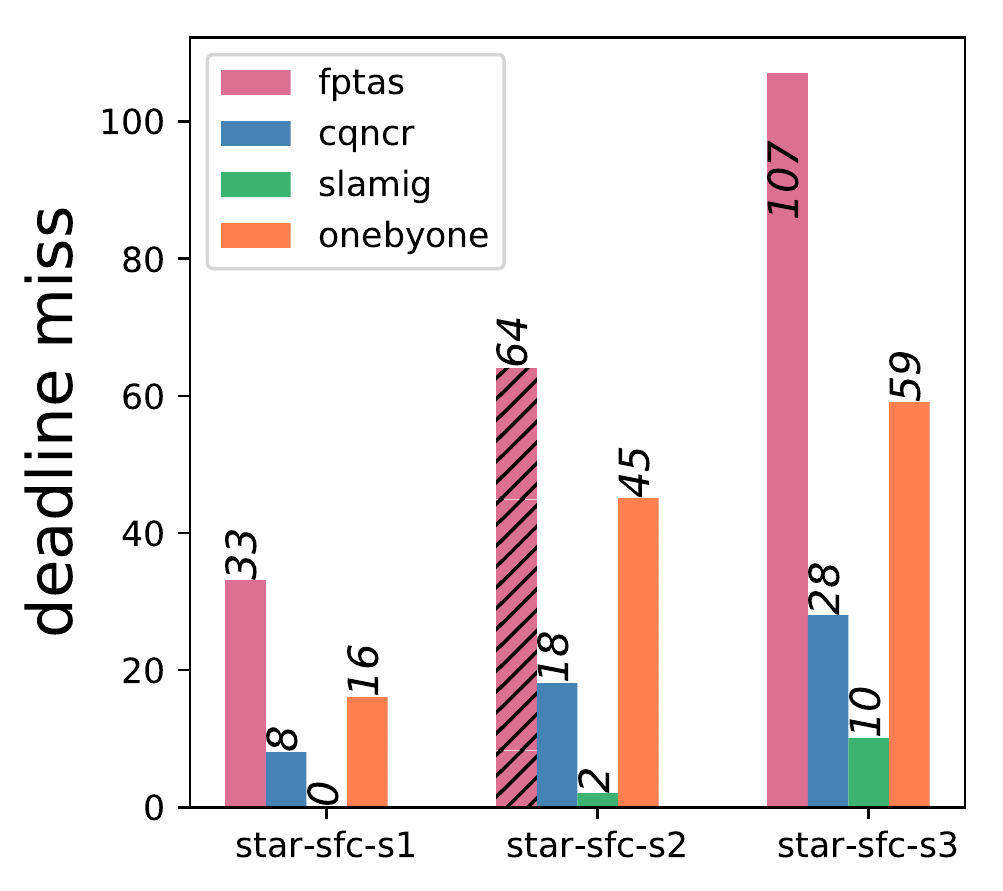}
		\subcaption{total missing deadline}
		\label{fig: deadline2}
	\end{subfigure}\hfil
	\caption{Deadline-related experiments in inter-datacenter}
	\label{fig: deadline}
\end{figure}

In this section, we evaluate and analyze the performance of different multiple migration plans under various urgency and priorities. In the experiment \textit{star-sfc}, we evaluated the deadline awareness in the remaining scheduling window and the number of total missing deadlines. In Table \ref{tb: deadline-simulation}, as shown in the QoS-aware experiment, instances in \textit{star-to-slave} have large delays due to the burst workloads, so the deadlines are tight. Meanwhile, the deadline for migration VNFs in \textit{sfc} with sufficient bandwidth is larger. The dirty page factor is 0.02 for all instances in this experiment.

Fig. \ref{fig: deadline} illustrates the results of the remaining scheduling window and the total missing deadlines. By ignoring the nature of migrations with various urgency and priorities, the two algorithms (FTPAS and CQNCR) as a comparison have unacceptable performance in terms of the remaining scheduling windows and the number of migration deadline violations. The average number of remaining scheduling window of FPTAS is negative due to the large execution time by allowing insufficient migration bandwidth. In all three scenarios, SLAMIG has the most remaining scheduling window, which can reduce SLA violations and guarantee the QoS during the migration with different priorities. Compared with FPTAS, CQNCR, and the baseline, FPTAS reduces the deadline violations by $100\%$, $96.875\%$/$88.89\%$/$95.56\%$, and $90.65\%$/$64.29\%$/$83.08\%$.

\paragraph{Summary}
By comprehensively considering the scheduling window, execution time, and the impact of one migration, SLAMIG can efficiently reduce the deadline missing while achieving the optimal migration performance. As a result, the total number of SLO violations can be minimized. Due to the flexibility of SLAMIG, one can also change the weight function to further reduce the migration deadline violations by trading off the performance of total migration time.

\subsubsection{Energy Consumption}
\begin{figure}[th]
	\centering
	\begin{subfigure}{0.33\linewidth}
		\includegraphics[width=\linewidth]{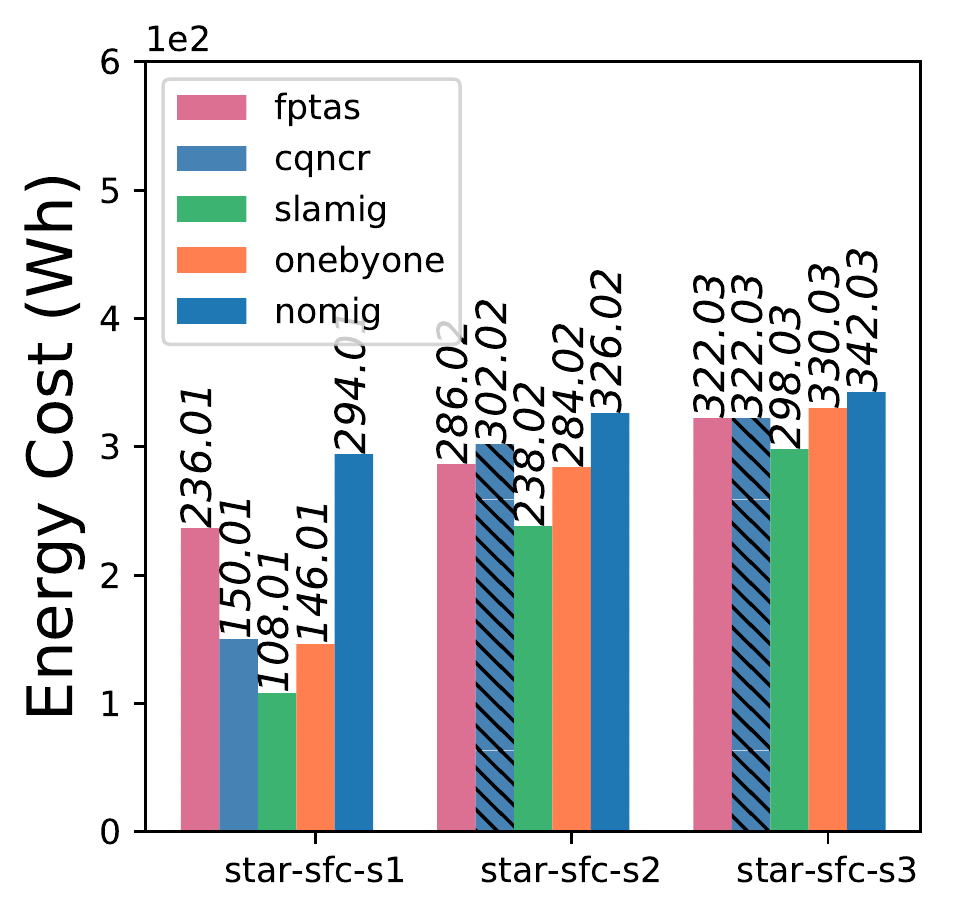}
		\subcaption{host energy in star-sfc}
		\label{fig: energy1}
	\end{subfigure}\hfil
	\begin{subfigure}{0.33\linewidth}
		\includegraphics[width=\linewidth]{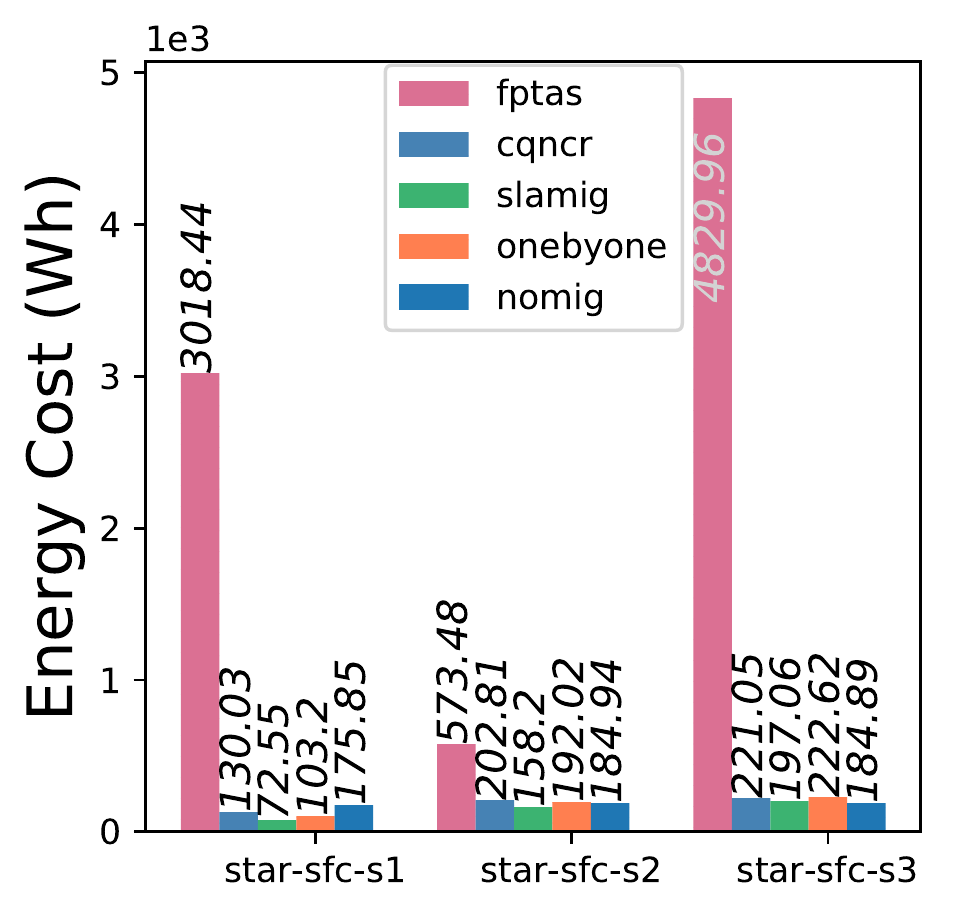}
		\subcaption{switch energy in star-sfc}
		\label{fig: energy2}
	\end{subfigure}\hfil \\
\begin{subfigure}{0.33\linewidth}
	\includegraphics[width=\linewidth]{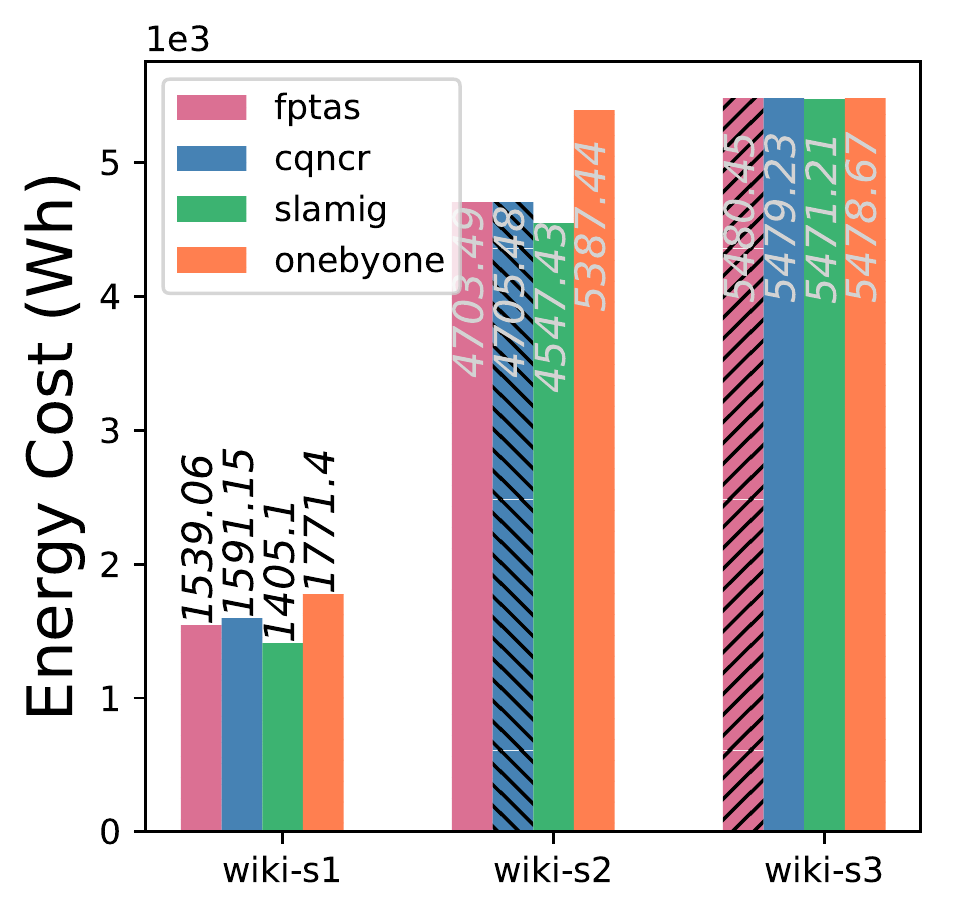}
	\subcaption{host energy in wiki}
	\label{fig: energy3}
\end{subfigure}\hfil
\begin{subfigure}{0.33\linewidth}
	\includegraphics[width=\linewidth]{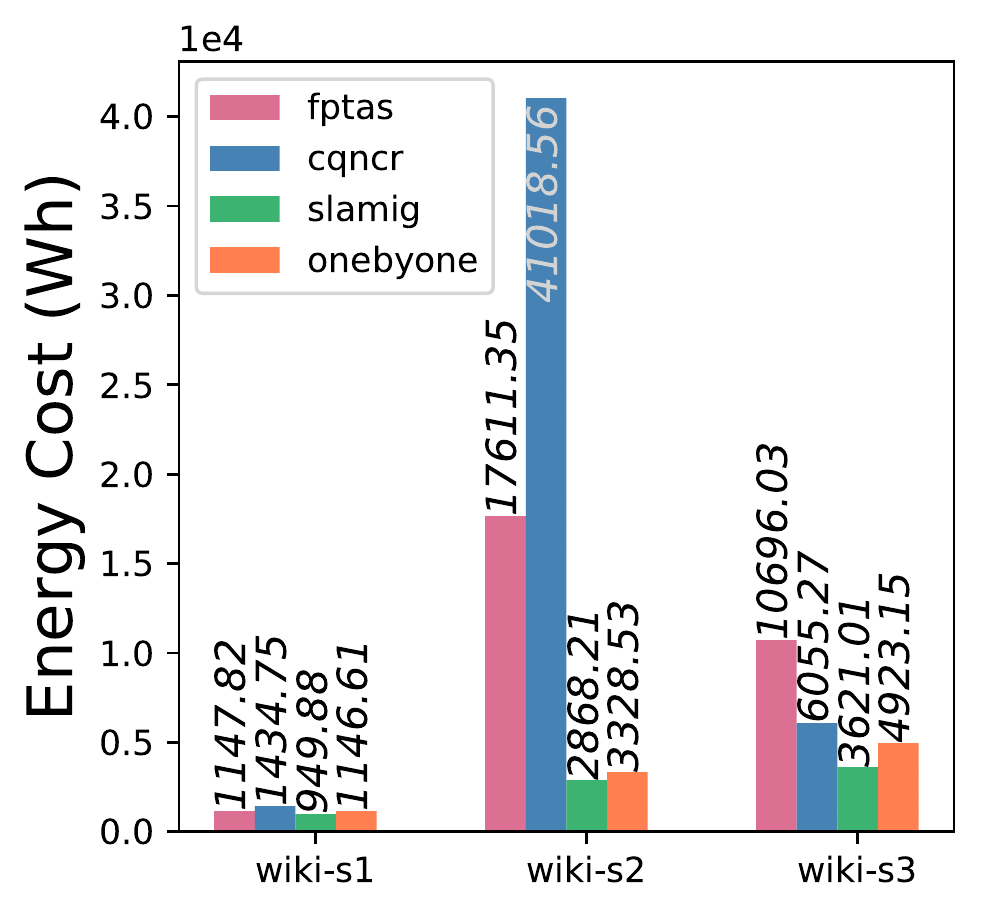}
	\subcaption{switch energy in wiki}
	\label{fig: energy4}
\end{subfigure}\hfil
	\caption{Energy consumption in hosts and switches}
	\label{fig: energy}
\end{figure}

In this section, we evaluate and analyze how different multiple migration plans can affect the energy consumption of hosts and switches. Switch \cite{wang2012carpo} and Host \cite{pelley2009understanding} power models are used to calculate the overheads of multiple live migrations in the data centers. Fig. \ref{fig: energy} shows the power consumption of host and switch in experiment \textit{star-sfc} and \textit{wiki}. As fewer hosts are involved after the consolidation, earlier migration convergence can reduce the host power consumption. In the experiment of \textit{star-sfc}, SLAMIG reduces the host power consumption by $63.26\%$, $26.99\%$, and $12.86\%$ compared to the non-migration and reduces by $26.03\%$, $16.20\%$, and $7.45\%$ compared to the second-best results. We also observed similar results of host energy consumption in wiki-s1 and wiki-s2 scenarios. In wiki-s3, due to involved hosts are consistent after migrations, there are only ignorable variances of host energy consumption among different algorithms. 

For the power consumption in networking resources (switches), the main contribution comes from the elephant flows of migrations from source to destination hosts. Another contribution comes from the application communications, where requests are sent between different physical hosts. In Fig. \ref{fig: energy2}, the networking energy consumption of FPTAS is much larger than other algorithms because it allows small bandwidth allocation to maximize the global migration network transmission rate. Our proposed approach is 29.70\%, 17.61\%, 10.85\% less than the second-best result. Fig. \ref{fig: energy4} indicates that SLAMIG also comsumes the least energy during the multiple migrations in \textit{wiki} experiment. Compared with the sequential scheduling, SLAMIG reduces by 17.16\%, 13.83\%, and 26.45\%. As mentioned, although the total migration time of FPTAS is smaller in wiki-s1, it costs 197.94Wh more than the SLAMIG. Therefore, the average migration execution time is also related to the migration overhead of energy consumption.

\paragraph{Summary}
Although smaller total migration time can reduce the total energy consumption due to consolidation, maintaining the average execution time is critical to the network power consumption. Due to the heavy usage of network resources, the switches consume a lot of energy during the migration. Even though consolidation and dynamic switching off the switches and hosts can help data centers save energy, migrating high-dirty-rate instances will increase the energy consumption of switches. Therefore, multiple migration tasks must be carefully planned based on the migrating candidates, sources and destination hosts. Dynamic resource management policies also need to consider the trade-off between the optimal allocation and migration energy overheads.

\section{Related Work}
\begin{table}[htbp]
	\centering
	\caption{Comparison of approaches on multiple migration planning and scheduling}
	\resizebox{\linewidth}{!}{
		\begin{tabular}{|l||l|l|l|l|l|l|l|}
			\hline
			App- 					& deadline		&  QoS 		&	energy	&	concurrent mig.	& online mig.	  &	multipath	 &	objectives   \\
			roach					& awareness		&  awareness&	consumption	& scheduling	& scheduler       &	routing		& 				 \\
			\hline
			\cite{checconi2009real}&	\checkmark	&	x		&	x		&	x		        &		x	    &	x			&	reduce the dirty memory transmission	\\
			\cite{ghorbani2012}		&	x			&\checkmark	&	x		&		x			&	x			&	x			&	sequence for loop-free and bandwidth constraints \\
			\cite{mann2012},\cite{xu2014}&x			&\checkmark	&	x		&		-			&	-			&	-			&	select migrating VMs to minimize interference\\
			\cite{bari2014cqncr}   &	x	        &\checkmark &	x		&	\checkmark		&  x	        &	x		    &	total mig. time and downtime with reserved bandwidth sharing	\\
			\cite{tsakalozos2017live}&	\checkmark	&	x		&	x		&	-		        &  \checkmark	&	x		    &	converge migration tasks before deadline	\\
			\cite{wang2017virtual}  &	x	        &	x		&	x		&	\checkmark		&  x	        &	\checkmark	&	total mig. time and downtime with free bandwidth sharing	\\
			SLAMIG  				&	\checkmark	&\checkmark	&\checkmark &	\checkmark		&  \checkmark	&	\checkmark	&	total mig. time, downtime, avg. exe. time, and transferred data	\\
			\hline
		\end{tabular}
	}
\label{tb: realted-work}
\end{table}

Akoush et al. \cite{akoush2010} explored the important parameters, link bandwidth and page dirty rate, that affect migration performance. They conducted experiments on migration performance under various workloads and proposed two simulation models based on the assumption of average memory dirty rate and history-based dirty rate of VM to predict migration performance. There are some works on the VM migration selector to minimize the overall cost and reduce interference. Remedy \cite{mann2012} relied on the SDN controller to monitor the state of the data center network and predict the cost of VM migration. The VM migration controller of heuristic destination selector minimizes the migration impact on the network by considering the cost of migration, the available bandwidth for migration, and the network balance achieved after migration. iAware \cite{xu2014} proposed a simple and light-weight interface-aware VM live migration strategy. It jointly estimates and minimizes the overall performance overhead of both migration interference and VM co-location interference with respect to I/O, CPU, and memory resources during and after migration. 

There are Few studies related to the (soft) real-time issue in live VM migration. These studies mainly focused on how to reduce the execution time of a single live migration. Tsakalozos et al. \cite{tsakalozos2017live} studied the live VM migration with time-constraints in the sharing-nothing IaaS-Clouds, where the cloud operator can assign specific scheduling windows for each migration task. For alleviating the SLA violations, they proposed a migration broker to monitor and limit the resource consumption, that is, to reduce the dirty page rate to force certain migrations to converge on time. By investigating the computing and network resources used by single live migration, Checconi et al. \cite{checconi2009real} presented a method to delay the frequent page dirtying in order to reduce the execution time and downtime of a single live migration.

Furthermore, there are several works focus on optimizing multiple live VM migration planning.
Ghorbani et al. \cite{ghorbani2012} proposed a simple one-by-one heuristic VM migration planning, which did not consider parallel VM migration through different network paths. 
Sun et al. \cite{sun2016} explore the optimal planning for multiple VM migrations by mixing pre-copy and post-copy migration. Based on the fact of application network traffic direction characteristic, it maximizes the available bandwidth to improve serial and parallel migrations. Similarly, Deshpande et al. \cite{deshpande2017} improved the live migration performance by considering pre-copy or post-copy migration based on the application traffic direction.
CQNCR \cite{bari2014cqncr} focuses on the multiple VM migration planning in one data center environment by considering the available bandwidth and network traffic cost after migration. They modeled the multiple VM migration planning based on a discrete-time model as a Mixed-Integer Programming (MIP) problem. A heuristic migration grouping algorithm by setting the group start time based on the prediction model is proposed. However, because there are different combinations of migration grouping, grouping and weighting the migration groups directly can lead to performance degradation of the total migration time.
Without considering the connectivity between VMs and the change of bandwidth, FPTAS \cite{wang2017virtual} simplifies the problem by maximizing the net transmission rate rather than minimizing the total migration time. In the context of SDN, the primary contribution compared to other research is the introduction of the multipath transmission when migrating VMs. As a MIP problem, they propose a fully polynomial-time approximation by further omitting certain variables. Table \ref{tb: realted-work} summarizes the comparison of live migration planning and scheduling methods for the objectives to be migrated, and whether the deadline of different migration tasks, QoS of applications, the energy consumption of hosts and switches, concurrent migration scheduling, and enables the multiple routing of migration flows and online scheduler to manage migration tasks are considered. The dash mark indicates the parameter of the work is not relevant.

\section{Summary and Conclusions}
Due to the limited computing and network resources as well as migration overheads, it is essential to intelligently schedule the migration tasks in data centers to achieve optimal migration performance, while mitigating the impacts of migration on cloud services and preventing SLO violations during the migration schedule. In this paper, we proposed a set of algorithms (SLAMIG) which includes concurrent migration grouping and the on-line migration scheduler. Instead of grouping migrations directly, SLAMIG can optimize the order of concurrent migration groups by sorting each migration based on complete dependency subgraphs. In addition to the dirty page rate,  extra bandwidth constraints can significantly improve the performance. The on-line migration scheduler can guarantee the concurrency and scheduling order of different migrations in a dynamic network environment. 

We argue that along with the total migration time, optimizing the average execution time, transferred data, and downtime are essential metrics to evaluate the multiple migration performance. The total migration time is more related to the time requirements (for example, migration deadlines and SLO violations), while the sum of execution time, transferred data, and service downtime are related to the actual overheads. By optimizing the total migration time, we can guarantee the SLA and dynamic performance requirements of cloud services. By optimizing the sum of execution time, transferred data, and downtime, we can guarantee the QoS of services and achieve more revenue as the cloud provider.
Experimental results show that SLAMIG can efficiently reduce the number of migration deadline missing and meanwhile achieve good migration performance in total migration time, average execution time, downtime, transferred data with acceptable algorithm runtime. Furthermore, the average execution time is an essential parameter to minimize the impact of multiple migration scheduling on the QoS of applications and energy consumption.

\section*{Acknowledgments}
This work is partially supported by an ARC Discovery Project and a Chinese Scholarship Council studentship. 
We thank Editor-in-Chief, Area Editor, and
reviewers for their valuable comments and suggestions that
helped in improving the paper significantly.

\section*{References}

\bibliography{ref-full}

\end{document}